\documentclass[journal,onecolumn]{IEEEtran}

\usepackage{color}
\usepackage{graphicx}
\usepackage{pstool}
\usepackage{overpic}
\usepackage[usestackEOL]{stackengine}
\usepackage{array,booktabs,multirow}
\usepackage{cancel}
\usepackage{bm}
\usepackage{mathbbol}
\usepackage{stfloats}
\usepackage{tabularx}
\usepackage{blindtext}
\usepackage{siunitx}
\usepackage[subrefformat=parens,labelformat=parens]{subfig}
\usepackage{mathtools}

\newcolumntype{N}{@{}m{0pt}@{}}
\newcommand{\htext}[1]{%
	\makebox[0pt]{\Centerstack{#1}}
}

\newcommand{\vtext}[1]{%
	\makebox[0pt]{\rotatebox[origin=c]{90}{\Centerstack{#1}}}
}

\newcommand{\tensor}[1]{\overline{\overline{#1}}}

\newcommand{\chia}[2]{\chi_{\text{#1}}^{#2}}

\newcommand{\chiat}[1]{\tensor{\chi}_{\text{#1}}}

\newcommand{\Ev}{\mathbf{E}}
\newcommand{\Hv}{\mathbf{H}}
\newcommand{\Dv}{\mathbf{D}}
\newcommand{\Bv}{\mathbf{B}}
\newcommand{\Pv}{\mathbf{P}}
\newcommand{\Mv}{\mathbf{M}}
\newcommand{\Qt}{\tensor{Q}}
\newcommand{\St}{\tensor{S}}
\newcommand{\Jv}{\mathbf{J}}

\newcommand{\zv}{\mathbf{\hat{z}}}

\newsavebox{\varmatrixbox}

\ExplSyntaxOn
\keys_define:nn {martin/varmatrix}
 {
  sep   .dim_set:N = \l_martin_varmatrix_sep_dim,
  delim .tl_set:N  = \l_martin_varmatrix_delim_tl,
  size  .tl_set:N  = \l_martin_varmatrix_size_tl,
  sep   .initial:n = 0.7\arraycolsep,
  size  .initial:n = \small,
 }

\NewDocumentEnvironment{varmatrix}{O{}}
 {
  \keys_set:nn {martin/varmatrix} { #1 }
  \begin{lrbox}{\varmatrixbox}
  \l_martin_varmatrix_size_tl
  \setlength{\arraycolsep}{\l_martin_varmatrix_sep_dim}
  $\begin{\l_martin_varmatrix_delim_tl matrix}
 }
 {
  \end{\l_martin_varmatrix_delim_tl matrix}$
  \end{lrbox}
  \vcenter{\box\varmatrixbox}
 }
\ExplSyntaxOff

\begin{document}

\title{Quadrupolar susceptibility modeling of substrated metasurfaces with application to the generalized Brewster effect}

\author{Ville Tiukuvaara, Olivier J. F. Martin, and Karim Achouri\\
Nanophotonics and Metrology Laboratory, Swiss Federal Institute of Technology Lausanne (EPFL), EPFL-STI-NAM, Station 11, CH-1015 Lausanne, Switzerland\\
email: karim.achouri@epfl.ch}

\maketitle



\begin{abstract}
We derive generalized sheet transition conditions (GSTCs) including dipoles and quadrupoles, using generalized functions (distributions). This derivation verifies that the GSTCs are valid for metasurfaces in non-homogeneous environments, such as for practical metasurfaces fabricated on a substrate. The inclusion of quadrupoles and modeling of spatial dispersion provides additional hyper-susceptibility components which serve as degrees of freedom for wave transformations. We leverage them to demonstrate a generalized Brewster effect with multiple angles of incidence at which reflection is suppressed, along with an ``anti-Brewster'' effect where transmission is suppressed.
\end{abstract}

\section{Introduction}

The study of metamaterials, and metasurfaces in particular, has reached a level of maturity such that recent works present increasingly elaborate applications. While at first, metasurfaces were used to provide simple wave transformations \cite{glybovskiMetasurfacesMicrowavesVisible2016,chenReviewMetasurfacesPhysics2016} and flat optics \cite{yuFlatOpticsDesigner2014}, they have now been used for sophisticated holography \cite{genevetHolographicOpticalMetasurfaces2015} and recently for computation and signal processing \cite{abdollahramezaniMetaopticsSpatialOptical2020,xueHighNAOpticalEdge2021,chenOnchipOpticalSpatialdomain2021,babaeeParallelAnalogComputing2021,momeniReciprocalMetasurfacesOnaxis2021}. This last application -- where transfer functions are implemented in the Fourier domain -- has required intricate design of meta-atoms to achieve control of the angular scattering response \cite{momeniReciprocalMetasurfacesOnaxis2021,momeniAsymmetricMetalDielectricMetacylinders2021,achouriAngularScatteringProperties2020}.

To aid in the design of metasurfaces for these applications, several modeling techniques have established themselves \cite{liuGeneralizedBoundaryConditions2019}. One popular approach is to model a metasurface as an impedance sheet which supports electric and magnetic currents \cite{liuGeneralizedBoundaryConditions2019,wangIndependentControlMultiple2020,montiSurfaceImpedanceModeling2020}. This induces boundary conditions on the tangential parts of the electric and magnetic fields. However, the impedances do not provide characteristic parameters to represent the metasurface since they depend on the incident fields \cite{tiukuvaaraSurfaceSusceptibilitiesCharacteristic2022}. A second approach is to determine the polarizability of an isolated meta-atom and account for its coupling through the array to other meta-atoms using Green's functions in the so-called T-matrix approach \cite{rahimzadeganComprehensiveMultipolarTheory}. This provides insights into the multipole moments which are present, and how they couple together. However, it does not serve as a boundary condition; rather, it provides the scattered fields when the incident field is specified. The last popular approach is the use of surface susceptibilities, which represent the metasurface as a zero-thickness sheet of multipole moment densities \cite{dehmollaianComparisonTensorBoundary2019,zaluskiAnalyticalExperimentalCharacterization2016,hollowayDiscussionInterpretationCharacterization2009,hollowayHomogenizationTechniqueObtaining2016,hollowayCharacterizingMetasurfacesMetafilms2011}. Given these moments along a surface, generalized sheet transition conditions (GSTCs) provide boundary conditions on the fields adjacent to the surface \cite{kuesterAveragedTransitionConditions2003}. These have been used to design metasurfaces \cite{achouriGeneralMetasurfaceSynthesis2015} and also implemented in numerical methods to greatly decrease the computational resources needed for their analysis \cite{smyFDTDSimulationDispersive2020,smyIEGSTCMetasurfaceField2021,smyPartEigenfunctionExpansion2022}.

Recently, it has become evident that susceptibility modeling, which was previously limited to the dipolar regime, should include spatial dispersion (nonlocality)~\cite{bernalarangoUnderpinningHybridizationIntuition2014}. This was analyzed with ``angle-dispersive'' dipolar susceptiblities in \cite{nizerrahmeierPartSpatiallyDispersive2022} while we considered higher-order multipoles in \cite{achouriExtensionLorentzReciprocity2021,achouriMultipolarModelingSpatially2022}. These considerations are especially true for optical metasurfaces, which generally have large meta-atoms with dimensions that approach the wavelength. Using GSTCs that were generalized to include quadrupoles, we demonstrated an improvement in the modeling accuracy~\cite{achouriMultipolarModelingSpatially2022}. In addition to improving the accuracy, the additional susceptibility components provide additional degrees of freedom for designing metasurfaces. 

However, the derivations in~\cite{achouriMultipolarModelingSpatially2022} are limited since they assume the media below and above the metasurface to be identical. Thus, it is not a priori obvious whether they would rigorously apply to practical metasurfaces which are usually fabricated on a substrate. In this work, we overcome the limitation of~\cite{achouriMultipolarModelingSpatially2022} by deriving the GSTCs, but using a different approach based on distributions (generalized functions)~\cite{richardsTheoryDistributionsNontechnical1990}, inspired by the work of Idemen~\cite{mithatDiscontinuitiesElectromagneticField2011}. Ultimately, our derivation produces GSTCs identical to those in~\cite{achouriMultipolarModelingSpatially2022}, which demonstrates that the latter can indeed be used in the presence of a substrate.

To demonstrate the utility of these GSTCs, we demonstrate the full control of the Brewster angle, where reflection at a dielectric interface is suppressed at a particular angle. By placing a metasurface at the dielectric interface, it is possible to tune the Brewster angle, as shown in \cite{paniagua-dominguezGeneralizedBrewsterEffect2016,lavigneGeneralizedBrewsterEffect2021}. We now leverage the higher-order susceptibility components to show that the additional degrees of freedom allow for further control, such as multiple Brewster angles, and suppression of transmission at particular angles---which we call ``anti-Brewster'' angles.

This paper is outlined as follows. First, we introduce generalized functions and derive the GSTCs in Section~\ref{sec:multipolar-gstcs}. Next, Section~\ref{sec:nonlocal} presents considerations to enforce the physicality of the analysis: spatial dispersion, properties of the moment tensors, and spatial symmetries of meta-atoms. Then, several examples of controlling the Brewster and ``anti-Brewster'' angles are presented in Section~\ref{sec:brewster}. Finally, we conclude in Section~\ref{sec:conclusion}.


\section{GSTCs with Quadrupoles}
\label{sec:multipolar-gstcs}
In this section, we will generalize the GSTCs to account for quadrupolar moments. Such a derivation was performed in \cite{achouriMultipolarModelingSpatially2022}, but with a caveat: the derivation assumed the bulk media adjacent to the metasurface to be homogeneous, and identical on both sides. This limitation arose from the use of the vector potential of the surface currents. We will overcome this limitation using an alternative derivation which represents the fields using distributions (generalized functions), following the approach taken by Idemen \cite{mithatDiscontinuitiesElectromagneticField2011}. Using this approach, the bulk material properties can be arbitrary as they are embedded in the definitions of the fields.

\begin{figure}[h]
    \centering
    \begin{overpic}[grid=false,trim={0cm 0cm 0cm 0cm},clip,tics=5]{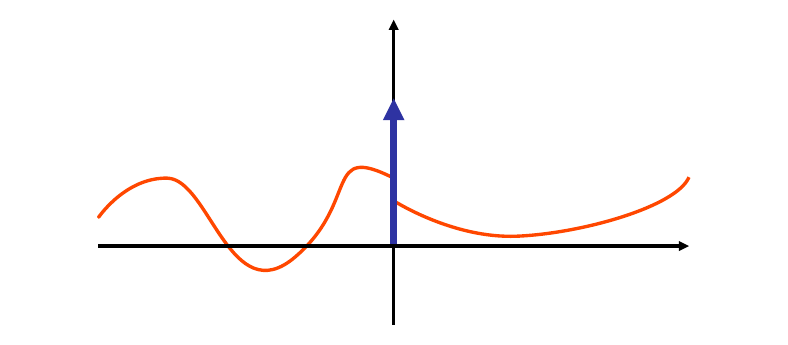}
        \put(30,22){\htext{\small{}\color{red}$\left\{\Lambda(z)\right\}$}}
        \put(66,27){\htext{\small{}\color{blue}$\sum_{k=0}^{\infty}\Lambda_k\delta^{(k)}(z)$}}
        \put(50,43){\htext{\small{}$\Lambda(z)$}}
        \put(90,11.5){\htext{\small{}$z$}}
    \end{overpic}
    \caption{A arbitrary discontinuous function $\Lambda(z)$ can be divided into a continuous part $\{\Lambda(z)\}$ and the discontinuity expressed as a summation of Dirac delta functions.}
    \label{fig:discontinuity}
\end{figure}
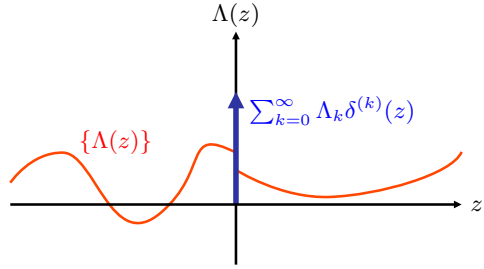

Distributions are ideal for modeling metasurfaces as zero-thickness discontinuities, since they formalize the notion of an ``impulse function''. For example, the electric polarization of a flat metasurface in the $xy$ plane may be expressed as $\mathbf{P}(x,y,z)=\mathbf{P}'(x,y)\delta(z)$, where $\delta(z)$ is the Dirac delta distribution, which in turn is rigorously defined using test functions \cite{mithatDiscontinuitiesElectromagneticField2011}. More generally, any field quantity $\Lambda$ can be decomposed into a continuous part and a discontinuous part:
\begin{align}
    \Lambda(z)=\{\Lambda(z)\}
    +\sum_{k=0}^\infty\Lambda_k\delta^{(k)}(z) \,,
    \label{eq:generalized-function}
\end{align}
where $\{\Lambda(z)\}$ represents the continuous part of $\Lambda(z)$ and a summation of the Dirac distribution and its derivatives is used to represent the discontinuity, as in Fig.~\ref{fig:discontinuity}. By interpreting Maxwell's equations with all field quantities as distributions, discontinuities in the fields are acceptable and treated rigorously, one arrives at a new set of equations called the \textit{universal boundary conditions} \cite{mithatDiscontinuitiesElectromagneticField2011,idemenBoundaryConditionsElectromagnetic1987}.

The starting point to derive the GSTCs are the universal boundary conditions and the following relations for the electric and magnetic flux densities for media with multipolar responses \cite{achouriExtensionLorentzReciprocity2021,simovskiCompositeMediaWeak2018}:
\begin{subequations}
\begin{align}
    \mathbf{D}&=\epsilon_{0} \mathbf{E}+\mathbf{P}-\frac{1}{2} \overline{\overline{Q}} \cdot \nabla\label{Eq:GSTCDerivationFlux}\\
			\mathbf{B} &= \mu_{0}\left(\mathbf{H}+\mathbf{M} -\frac{1}{2} \overline{\overline{S}} \cdot \nabla\right) \,,
\end{align}\label{eq:DB}
\end{subequations}
where $\Pv$ is the electric dipole density, $\Qt$ is the electric quadrupolar moment density, $\Mv$ is the magnetic dipolar density, and $\St$ is the magnetic quadrupolar density. For a metasurface, each of the quadrupolar moments may be represented by a single sheet discontinuity (no continuous part and $k=0$ from \eqref{eq:generalized-function}) and so we re-write~\eqref{eq:DB} as
\begin{subequations}
\begin{align}
    \mathbf{D}&=\epsilon_{0} \mathbf{E}+ \{\mathbf{P}\} +\mathbf{P}_0\delta^{(0)}(z)-\frac{1}{2} \left[\overline{\overline{Q}}_0\delta(z)^{(0)}\right] \cdot \nabla\label{Eq:GSTCDerivationFluxDistribution}\\
			\mathbf{B} &= \mu_{0}\left(\mathbf{H}+ \{\mathbf{M}\}+\mathbf{M}_0\delta^{(0)}(z) -\frac{1}{2} \left[\overline{\overline{S}}_0\delta^{(0)}(z)\right] \cdot \nabla\right) \,,
\end{align}\label{eq:DB-distribution}
\end{subequations}
where the bulk polarization (with possibly different media on the two sides of the metasurface) is embedded within $\{\mathbf{P}\}$ and $\{\mathbf{M}\}$ as well as within the fields $\Ev, \Hv, \Dv$ and $\Bv$. Then, by simplifying~\eqref{eq:DB-distribution} and substituting it into the universal boundary conditions as shown in the supplementary information, one arrives at the following GSTCs: 
\begin{subequations}
    \begin{multline}
		\qquad\qquad\qquad\qquad\qquad\mathbf{z} \times \Delta\mathbf{E}=
		-j\omega\mu_0\Mv_{t}
		+\frac{k_0^2}{2\epsilon_0}\zv\times\left(\Qt\cdot\zv\right)
		\\-\frac{1}{\epsilon_0}\zv\times \nabla_t\left[
		P_{z}-\frac{1}{2}(\nabla_t\zv+\zv\nabla_t):\Qt
		\right]
		+\frac{j\omega\mu_0}{2}\left[
		\left(\St-S_{zz}\tensor{I}\right)\cdot\nabla_t
		\right]_t
		\end{multline}
		\begin{multline}
		\qquad\qquad\qquad\qquad\qquad\mathbf{z} \times\Delta \mathbf{H}=
		j\omega\Pv_{t}
		+\frac{k_0^2}{2} \zv\times\left(\St\cdot\zv\right)
		\\-\zv\times\nabla_t\left[M_{z}-\frac{1}{2}(\nabla_t\zv+\zv\nabla_t):\St\right]
		-\frac{j\omega}{2}
		\left[
		\left(\Qt-Q_{zz}\tensor{I}\right)\cdot\nabla_t
		\right]_t \,,
		\end{multline}
where $\tensor{I}$ is the identity matrix and the $t$ subscript indicates the tangential components. These are in agreement with those derived independently in \cite{achouriMultipolarModelingSpatially2022}; however, our derivation proves that the GSTCs are independent of the media on either side of the metasurface, since there is no restrictions on the bulk moments, $\{\mathbf{P}\}$ and $\{\mathbf{M}\}$, in~\eqref{eq:DB-distribution}. Nevertheless, the information regarding the material parameters of these media remains present in these equations as it is embedded within the definition of the fields that interact with the metasurface. Additionally, our derivation provides boundary conditions on the normal components of the fields, which have not yet been shown in the literature:
		\begin{gather}
		\mathbf{z} \cdot \Delta\mathbf{D}
		=-\nabla_t\cdot
		\left(
		\Pv_{t}
		- \frac{j\omega\mu_0\epsilon_0}{2}\zv\times\left(\St\cdot\zv\right)
		-
			\left(\Qt-Q_{zz}\tensor{I}\right)\cdot\nabla_t
		\right)\\
		\mathbf{z} \cdot \Delta\mathbf{B}
		= -\nabla_t\cdot\mu_0\left(\Mv_{t}
		+\frac{j\omega}{2}\zv\times\left(\Qt\cdot\zv\right)- 
		\left(\St-S_{zz}\tensor{I}\right)\cdot\nabla_t
		\right) \,.
		\end{gather}\label{eq:GSTCs}
\end{subequations}
Naturally, these simplify to the conventional textbook boundary conditions for a dielectric interface when there are no surface polarization moments, which can be easily verified by setting $\Pv=\Mv=\Qt=\St=0$.

\section{Non-Local Considerations}\label{sec:nonlocal}
The GSTCs \eqref{eq:GSTCs} provide boundary conditions, but do not model how the polarization densities arise. These can be captured using multipolar susceptibilities. In this section, we formulate constitutive relations, and discuss some considerations that must be taken when the metasurface is placed in a non-uniform environment.

\subsection{Acting Fields}
The fields which excite the meta-atoms are defined as the average of the adjacent fields on  either side of the metasurface. As in \cite{albooyehElectromagneticCharacterizationBianisotropic2016,achouriMultipolarModelingSpatially2022,achouriElectromagneticMetasurfacesTheory2021}, these average fields are usually defined as 
\begin{subequations}
    \begin{gather}
        \Ev_\text{av} = \frac{1}{2}\left.\left(\Ev_\text{i}+\Ev_\text{r}+\Ev_\text{t}\right)\right|_{z=0}\\
        \Hv_\text{av} = \frac{1}{2}\left.\left(\Hv_\text{i}+\Hv_\text{r}+\Hv_\text{t}\right)\right|_{z=0} \,.
    \end{gather}\label{eq:average-fields}
\end{subequations}
This works well if the metasurface is freestanding, but is inappropriate if the metasurface is placed between two different media. Considering a thin slab placed between two different media, it can be shown (see the supplementary information or \cite{kuesterElectromagneticBoundaryProblems2015}) that the normal components should be defined using the flux densities, $\epsilon E_{z}$ and $\mu H_{z}$, which remain continuous at a dielectric interface, such that the acting fields are
\begin{subequations}
    \begin{gather}
        \Ev_{\text{av}} =   \Ev_\text{av,t} +\frac{1}{2}\left.\left(\epsilon_1 E_{\text{i},z}+\epsilon_1 E_{\text{r},z}+\epsilon_2 E_{\text{t},z}\right)\right|_{z=0}\zv,\\
        \Hv_{\text{av}} =  \Hv_\text{av,t} +\frac{1}{2}\left.\left(\mu_1 H_{\text{i},z}+\mu_1 H_{\text{r},z}+\mu_2 H_{\text{t},z}\right)\right|_{z=0}\zv,
    \end{gather}\label{eq:acting-fields-fix}
\end{subequations}
where $\epsilon_1,~\mu_1$ and $\epsilon_2,~\mu_2$ correspond to the material parameters of the media at the top and the bottom sides of the metasurface, respectively.



\subsection{Spatial Dispersion}
Given the acting fields \eqref{eq:acting-fields-fix}, the constitutive relations can be written out. If the metasurface response may be described in terms of only dipolar components (e.g., when the metasurface unit cell is much smaller than the wavelength so that higher-order multipolar components are negligible), the induced moments are related using \cite{achouriGeneralMetasurfaceSynthesis2015,achouriMultipolarModelingSpatially2022}
\begin{subequations}
    \begin{align}
        \begin{bmatrix}
        \Pv \\ \Mv
        \end{bmatrix}
        =
        \begin{bmatrix}
        \epsilon_0\chiat{ee} & c_0^{-1}\chiat{em}\\
        \eta_0^{-1}\chiat{me} & \chiat{mm}\\
        \end{bmatrix}\cdot
        \begin{bmatrix}
        \Ev_\text{av}\\
        \Hv_\text{av}
        \end{bmatrix},
    \end{align}\label{eq:constitutive-local}
\end{subequations}
where the constants (i.e. $\epsilon_0$, $\eta_0$) are selected so the \emph{surface} susceptibilities have the unit of length~\cite{achouriElectromagneticMetasurfacesTheory2021}. However, when the metasurface unit cell becomes large, higher-order multipoles must be considered, starting with the quadrupoles. Including these adds a plethora of additional hyper-susceptibility components \cite{achouriMultipolarModelingSpatially2022}:



\begin{gather}
    \begin{bmatrix}
		P_i\\
		M_i\\
		Q_{il}\\
		S_{il}
	\end{bmatrix}=
    \begin{bmatrix}
		\epsilon_0\chi_{\text{ee}}^{ij} & \frac{1}{c_0}\chi_{\text{em}}^{ij} & \frac{\epsilon_0}{2k_0}\chi_{\text{ee}}^{'ijk} & \frac{1}{2c_0k_0}\chi_{\text{em}}^{'ijk}\\
		\frac{1}{\eta_0}\chi_{\text{me}}^{ij} & \chi_{\text{mm}}^{ij} & \frac{1}{2\eta_0k_0}\chi_{\text{me}}^{'ijk} & \frac{1}{2k_0}\chi_{\text{mm}}^{'ijk}\\
		\frac{\epsilon_0}{k_0}Q_{\text{ee}}^{ilj} & \frac{1}{c_0k_0}Q_{\text{em}}^{ilj} & \frac{\epsilon_0}{2k_0^2}Q_{\text{ee}}^{'iljk} & \frac{1}{2c_0k_0^2}Q_{\text{em}}^{'iljk}\\
		\frac{1}{\eta_0k_0}S_{\text{me}}^{ilj} & \frac{1}{k_0}S_{\text{mm}}^{ilj} & \frac{1}{2\eta_0k_0^2 }S_{\text{me}}^{'iljk} & \frac{1}{2k_0^2}S_{\text{mm}}^{'iljk}
	\end{bmatrix}
	\cdot
	\begin{bmatrix}
		E_{\text{av},j}\\
		H_{\text{av},j}\\
		\nabla_k E_{\text{av},j}\\
		\nabla_k H_{\text{av},j}
	\end{bmatrix} \,,\label{eq:spatial-dispersion}
\end{gather}
where in addition to quadrupolar susceptibilities which depend on the fields directly (e.g. $Q_{il}\propto Q_\text{ee}^{ilj}E_{\text{av},j}$), there are components which depend on the field gradients (e.g. $P_{i}\propto \chi_\text{ee}^{'ijk}\nabla_k E_{\text{av},j}$). This \textit{spatial dispersion} or \emph{nonlocality} is necessitated by reciprocity, which connects certain terms together (e.g. $Q_\text{ee}^{ilj}=\chi_\text{ee}^{'jli}$) \cite{achouriExtensionLorentzReciprocity2021}.
At this point, for simplicity but without loss of generality, we will consider TM-polarized plane-wave fields propagating in the $xz$ plane. Then, as shown in \cite{achouriMultipolarModelingSpatially2022}, \eqref{eq:spatial-dispersion} simplifies to
\begin{multline}
    \begin{bmatrix}
    P_x & P_z & M_y & Q_{xz} & Q_{xx}
     & Q_{zz} & S_{yz} & S_{yx}
    \end{bmatrix}^T
    \propto \\
    \tensor{X}\cdot
    \begin{bmatrix}
        E_{\text{av},x} & 
        E_{\text{av},z} & 
        H_{\text{av},y} & 
        \partial_x E_{\text{av},x} & 
        \partial_x E_{\text{av},z} + \partial_z E_{\text{av},x} & 
        \partial_z E_{\text{av},z} & 
        \partial_z H_{\text{av},y} & 
        \partial_x H_{\text{av},y}
    \end{bmatrix}^T \,,\label{eq:consitutive-general}
\end{multline}
where $\tensor{X}$ is the hypersusceptibilty matrix shown in Fig.~\ref{fig:chi-simplification}a. It is an $8\times 8$ matrix with 64 terms in general, but can be simplified by imposing conditions such as reciprocity and tracelessness, as we will do shortly. Before that, consider that it is non-sensical to include the derivatives $\partial_z$, since the values $E_{\text{av},x}$, $E_{\text{av},z}$ and $H_{\text{av},y}$ are independent of $z$, as defined in \eqref{eq:acting-fields-fix}. That is, they are functions of $x$ and $y$ only, and so the derivative ($\partial_z$) would be zero. However, we can overcome this issue by switching the order of operations; that is, by performing differentiation first, and then averaging. Furthermore, though the derivatives along $z$ are still problematic as they may be discontinuous, we can transform them into tangential derivatives using Maxwell's equations, as explained next.

First, consider Faraday's equation in either medium, $\nabla\times\Ev=j\omega\Bv$, which becomes $\partial_zE_x=\partial_xE_z-j\omega B_y$. Then, it follows that the spatial average of the derivative along $z$ of $E_x$ may be obtained as
\begin{subequations}
    \begin{align}
        \left.\left(\partial_zE_x\right)\right|_\text{av}
        &= \frac{1}{2}\left.\left(\partial_zE_{\text{i},x}+\partial_zE_{\text{r},x}+\partial_zE_{\text{t},x}\right)\right|_{z=0}\notag\\
        &= \partial_xE_{\text{av},z}-j\omega B_{\text{av},y}\notag\\
        &= \partial_xE_{\text{av},z}-\frac{j\omega}{2} \left.\left(\mu_1H_{\text{i},y}+\mu_1H_{\text{r},y}+\mu_2H_{\text{t},y}\right)\right|_{z=0} \,,
    \end{align}
where we have eliminated $\partial_z$. Next, consider Gauss' equation, $\nabla\cdot\Dv=0$ in either medium, that is, $\partial_zD_{z}=-\partial_xD_x$. Then,
    \begin{align}
        \left.\left(\partial_zE_z\right)\right|_\text{av}
        &= -\frac{1}{2}\left.\left(\epsilon_1\partial_xE_{\text{i},x}+\epsilon_1\partial_xE_{\text{r},x}+\epsilon_2\partial_xE_{\text{t},x}\right)\right|_{z=0} \,.
    \end{align}
Finally, consider Ampere's equation, $\nabla\times\Hv=j\omega\Dv$; that is, $\partial_zH_y=\partial_yH_z-j\omega D_x$.
    \begin{align}
        \left.\left(\partial_zH_y\right)\right|_\text{av}
        &= \frac{1}{2}\left.\left(\partial_zH_{\text{i},y}+\partial_zH_{\text{r},y}+\partial_zH_{\text{t},y}\right)\right|_{z=0}\notag\\
        &= \partial_yH_{\text{av},z}-j\omega D_{\text{av},x}\notag\\
        &= \partial_yH_{\text{av},z}-\frac{j\omega}{2} \left.\left(\epsilon_1E_{\text{i},x}+\epsilon_1E_{\text{r},x}+\epsilon_2E_{\text{t},x}\right)\right|_{z=0} \,.
    \end{align}\label{eq:elimiate-dz}
\end{subequations}
Now, with reference to \eqref{eq:elimiate-dz}, \eqref{eq:consitutive-general} is modified to
\begin{multline}
    \begin{bmatrix}
    P_x & P_z & M_y & Q_{xz} & Q_{xx}
    & Q_{zz} & S_{yz} & S_{yx}
    \end{bmatrix}^T
    \propto \\
    \tensor{X}\cdot
    \begin{bmatrix}
        E_{\text{av},x} & 
        E_{\text{av},z} & 
        H_{\text{av},y} & 
        \partial_x E_{\text{av},x} & 
        \partial_x E_{\text{av},z} + \left.\left(\partial_zE_x\right)\right|_\text{av} & 
        \left.\left(\partial_zE_z\right)\right|_\text{av} & 
        \left.\left(\partial_zH_y\right)\right|_\text{av} & 
        \partial_x H_{\text{av},y}
    \end{bmatrix}^T \,. \label{eq:consitutive-nodz}
\end{multline}

\begin{figure}[b!]
    \centering
    \newcommand{\PreserveBackslash}[1]{\let\temp=\\#1\let\\=\temp}
    \newcolumntype{C}[1]{>{\PreserveBackslash\centering}p{#1}}
    \begin{tabular}{C{0.48\textwidth}C{0.48\textwidth}}
        \scalebox{0.7}{\tiny$\begin{varmatrix}[delim=b,size=\normalsize,sep=1pt]
            \chi_\text{ee}^{xx}  & \chi_\text{ee}^{xz}   &  \chi_\text{em}^{xy}   &  \chi_\text{ee}^{'xxz}   &  \chi_\text{ee}^{'xxx}  &  \chi_\text{ee}^{'xzz}  &  \chi_\text{em}^{'xyx}  &  \chi_\text{em}^{'xyz}\\
			\chi_\text{ee}^{zx}  & \chi_\text{ee}^{zz}   &  \chi_\text{em}^{zy}   &  \chi_\text{ee}^{'zxz}   &  \chi_\text{ee}^{'zxx}  &  \chi_\text{ee}^{'zzz}  &  \chi_\text{em}^{'zyx}  &  \chi_\text{em}^{'zyz}  \\
			\chi_\text{me}^{yx}  & \chi_\text{me}^{yz}   &  \chi_\text{mm}^{yy}   &  \chi_\text{me}^{'yxz}   &  \chi_\text{me}^{'yxx}  &  \chi_\text{me}^{'yzz}  &  \chi_\text{mm}^{'yyx}  &  \chi_\text{mm}^{'yyz}  \\
			Q_\text{ee}^{xzx} & Q_\text{ee}^{xzz}  &  Q_\text{em}^{xzy}  &  Q_\text{ee}^{'xzxz}  &  Q_\text{ee}^{'xzxx}  &  Q_\text{ee}^{'xzzz}  &  Q_\text{em}^{'xzyx} &  Q_\text{em}^{'xzyz}  \\
			Q_\text{ee}^{xxx} & Q_\text{ee}^{xxz}  &  Q_\text{em}^{xxy}  &  Q_\text{ee}^{'xxxz}  &  Q_\text{ee}^{'xxxx}  &  Q_\text{ee}^{'xxzz}  &  Q_\text{em}^{'xxyx} &  Q_\text{em}^{'xxyz}  \\
			Q_\text{ee}^{zzx} & Q_\text{ee}^{zzz}  &  Q_\text{em}^{zzy}  &  Q_\text{ee}^{'zzxz}  &  Q_\text{ee}^{'zzxx}  &  Q_\text{ee}^{'zzzz}  &  Q_\text{em}^{'zzyx} &  Q_\text{em}^{'zzyz}  \\
			S_\text{me}^{yxx} & S_\text{me}^{yxz}  &  S_\text{mm}^{yxy}  &  S_\text{me}^{'yxxz}  &  S_\text{me}^{'yxxx}  &  S_\text{me}^{'yxzz}  &  S_\text{mm}^{'yxyx} &  S_\text{mm}^{'yxyz}  \\
			S_\text{me}^{yzx} & S_\text{me}^{yzz}  &  S_\text{mm}^{yzy}  &  S_\text{me}^{'yzxz}  &  S_\text{me}^{'yzxx}  &  S_\text{me}^{'yzzz}  &  S_\text{mm}^{'yzyx} &  S_\text{mm}^{'yzyz}  \\
		\end{varmatrix}$}\vspace{0.2cm}
		
		\small{}(a) Full susceptibility matrix ($8\times 8$)\vspace{0.2cm}
        &
        \scalebox{0.7}{\tiny$\begin{varmatrix}[delim=b,size=\normalsize,sep=1pt]
            \chi_\text{ee}^{xx}  & \chi_\text{ee}^{xz}   &  \chi_\text{em}^{xy}   &  \chi_\text{ee}^{'xxz}   &  \chi_\text{ee}^{'xxx}  &  -\chi_\text{ee}^{'xxx}  &  \chi_\text{em}^{'xyx}  &  \chi_\text{em}^{'xyz}\\
			\chi_\text{ee}^{xz}  & \chi_\text{ee}^{zz}   &  \chi_\text{em}^{zy}   &  \chi_\text{ee}^{'zxz}   &  \chi_\text{ee}^{'zxx}  &  -\chi_\text{ee}^{'zxx}  &  \chi_\text{em}^{'zyx}  &  \chi_\text{em}^{'zyz}  \\
			-\chi_\text{em}^{xy}  & -\chi_\text{em}^{zy}   &  \chi_\text{mm}^{yy}   &  \chi_\text{me}^{'yxz}   &  \chi_\text{me}^{'yxx}  &  -\chi_\text{me}^{'yxx}  &  \chi_\text{mm}^{'yyx}  &  \chi_\text{mm}^{'yyz}  \\
			\chi_\text{ee}^{'xxz} & \chi_\text{ee}^{'zxz}  &  -\chi_\text{me}^{'yxz}  &  Q_\text{ee}^{'xzxz}  &  Q_\text{ee}^{'xxxz}  &  -Q_\text{ee}^{'xxxz}  &  -S_\text{me}^{'yxzx} &  -S_\text{me}^{'yzzx}  \\
			\chi_\text{ee}^{'xxx} & \chi_\text{ee}^{zxx}  &  -\chi_\text{me}^{'yxx}  &  Q_\text{ee}^{'xxxz}  &  Q_\text{ee}^{'xxxx}  &  -Q_\text{ee}^{'xxxx}  &  -S_\text{me}^{'yxxx} &  -S_\text{me}^{'yzxx}  \\
			-\chi_\text{em}^{xyx} & -\chi_\text{em}^{zyx}  &  \chi_\text{mm}^{yyx}  &  S_\text{me}^{'yxzx}  &  S_\text{me}^{'yxxx}  &  -S_\text{me}^{'yxxx}  &  S_\text{mm}^{'yxyx} &  S_\text{mm}^{'yxyz}  \\
			-\chi_\text{em}^{xyz} & -\chi_\text{em}^{zyz}  &  \chi_\text{mm}^{yyz}  &  S_\text{me}^{'yzzx}  &  S_\text{me}^{'yzxx}  &  -S_\text{me}^{'yzxx}  &  S_\text{mm}^{'yxyz} &  S_\text{mm}^{'yzyz}  \\
		\end{varmatrix}$}\vspace{0.3cm}
		\small{}(b) Reciprocal \& traceless ($7\times 8$)
        \\
        \scalebox{0.7}{\tiny$\begin{varmatrix}[delim=b,size=\normalsize,sep=1pt]
			\chi_\text{ee}^{xx}  & 0   &  0   &  0   &  0 &  0  &  0  &  \chi_\text{em}^{'xyz}\\
			0  & \chi_\text{ee}^{zz}   &  0   &  0   &  0  &  0  &  0  &  0  \\
			0  & 0   &  \chi_\text{mm}^{yy}   &  \chi_\text{me}^{'yxz}   &  0  &  0  &  0  &  0  \\
			0 & 0  &  -\chi_\text{me}^{'yxz}  &  Q_\text{ee}^{'xzxz}  &  0 &  0  &  0 &  0  \\
			0 & 0  & 0  &  0  &  Q_\text{ee}^{'xxxx}  &  -Q_\text{ee}^{'xxxx}  &  0 &  0  \\
			0 & 0  &  0  &  0  &  0  &  0  &  S_\text{mm}^{'yxyx} &  0  \\
			-\chi_\text{em}^{xyz} & 0  &  0 &  0  &  0  &  0  &  0 &  S_\text{mm}^{'yzyz}  \\
		\end{varmatrix}$}\vspace{0.2cm}
		
		\small{}(c) All symmetries: $\sigma_x$, $\sigma_y$, $\sigma_z$, $C_{4z}$
		\vspace{0.2cm}
        &
        \scalebox{0.7}{\tiny$\begin{varmatrix}[delim=b,size=\normalsize,sep=1pt]
			\chi_\text{ee}^{xx}  & 0   &  \chi_\text{em}^{xy}   &  \chi_\text{ee}^{'xxz}   &  0 &  0  &  0  &  \chi_\text{em}^{'xyz}\\
			0  & \chi_\text{ee}^{zz}   &  0   &  0   &  \chi_\text{ee}^{'zxx}  &  -\chi_\text{ee}^{'zxx}  &  0  &  0  \\
			-\chi_\text{em}^{xy}  & 0   &  \chi_\text{mm}^{yy}   &  \chi_\text{me}^{'yxz}   &  0  &  0  &  0  &  \chi_\text{mm}^{'yyz}  \\
			\chi_\text{ee}^{'xxz} & 0  &  -\chi_\text{me}^{'yxz}  &  Q_\text{ee}^{'xzxz}  &  0 &  0  &  0 &  -S_\text{me}^{'yzzx}  \\
			0 & \chi_\text{ee}^{zxx}  & 0  &  0  &  Q_\text{ee}^{'xxxx}  &  -Q_\text{ee}^{'xxxx}  &  0 &  0  \\
			0 & 0  &  0  &  0  &  0  &  0  &  S_\text{mm}^{'yxyx} &  0  \\
			-\chi_\text{em}^{xyz} & 0  &  \chi_\text{mm}^{yyz}  &  S_\text{me}^{'yzzx}  &  0  &  0  &  0 &  S_\text{mm}^{'yzyz}  \\
		\end{varmatrix}$}\vspace{0.2cm}
		\small{}(d) All symmetries but $\sigma_z$
        \\
    \end{tabular}
    \caption{The general susceptibility matrix $\tensor{X}$, for TM-polarized fields propagating in the $xz$ plane, is shown in (a) and has 64 terms. By enforcing reciprocity and tracelenssness, this is reduced to 28 terms as in (b). Subsequently, spatial symmetries of the metasurface can be leveraged to further simplify the matrix, as shown for two examples in (c) and (d).}
    \label{fig:chi-simplification}
\end{figure}





\subsection{Tensor symmetries, tracelessness, and reciprocity}
Linear time-invariant metasurfaces that are not biased by a time-odd quantity (such as a magnetic field) are reciprocal and as such must satisfy reciprocity conditions~\cite{tretyakovAnalyticalModelingApplied2003,calozElectromagneticNonreciprocity2018}.
Furthermore, we note that all of the moments should be symmetrical; e.g. $Q_{ij}=Q_{ji}$ \cite{riccardiMultipolarExpansionsScattering2022}. These symmetry and reciprocity properties of the moments constrain the susceptibility terms such that they are related to one another as shown in \cite{achouriExtensionLorentzReciprocity2021}. Enforcing these relations reduces the 64 terms in Fig.~\ref{fig:chi-simplification}a to 36.

In addition to being symmetrical, the tensors should be traceless. Only if this is the case are the moments truly physically meaningful, and are called \text{irreducible} \cite{riccardiMultipolarExpansionsScattering2022}. In particular, note that $\sum_iQ_{ii}=0$, which implies $Q_{zz}=-Q_{xx}$ given that $Q_{yy}=0$ in our simplified problem. In contradiction to this condition, $Q_{zz}$ and $Q_{xx}$ are independent in \eqref{eq:consitutive-nodz}. 

To enforce tracelessness, note how the 6th row in Fig.~\ref{fig:chi-simplification}a should be the negative of the 5th row ($Q_{zz}=-Q_{xx}$), and can thus be eliminated. However, reciprocity must still be enforced: the 5th row and 5th column are related by reciprocity and tensor symmetries. Given the relationship between the 4th and 5th rows, reciprocity is maintained by rewriting the 5th column as the negative of the 4th column. Then, the 5th row can be eliminated and we arrive at the $7\times 8$ matrix in Fig.~\ref{fig:chi-simplification}b. This matrix ensures tensor symmetries, reciprocity, and traceless, and contains 28 unique terms. After eliminating $Q_{zz}$, \eqref{eq:consitutive-nodz} becomes
\begin{multline}
    \begin{bmatrix}
    P_x & P_z & M_y & Q_{xz} & Q_{xx} & S_{yz} & S_{yx}
    \end{bmatrix}^T
    \propto \\
    \tensor{X}\cdot
    \begin{bmatrix}
        E_{\text{av},x} & 
        E_{\text{av},z} & 
        H_{\text{av},y} & 
        \partial_x E_{\text{av},x} & 
        \partial_x E_{\text{av},z} + \left.\left(\partial_zE_x\right)\right|_\text{av} & 
        \left.\left(\partial_zE_z\right)\right|_\text{av} & 
        \left.\left(\partial_zH_y\right)\right|_\text{av} & 
        \partial_x H_{\text{av},y}
    \end{bmatrix}^T \,.\label{eq:consitutive-nodz-noQzz}
\end{multline}

\subsection{Spatial Symmetries}
Neumann's principle states that the material parameters of a system should exhibit the same symmetry properties as the physical structure they describe. This implies that if the considered physical structure (metasurface) is invariant under certain symmetry operations, then so should their material parameters (susceptibility tensors) \cite{achouriSymmetriesAngularScattering2019,achouriSpatialSymmetriesMultipolar2022}.

\begin{figure}
    \centering
    \begin{overpic}[width=\textwidth,grid=false,trim={0cm 0cm 0cm 0cm},clip,tics=2]{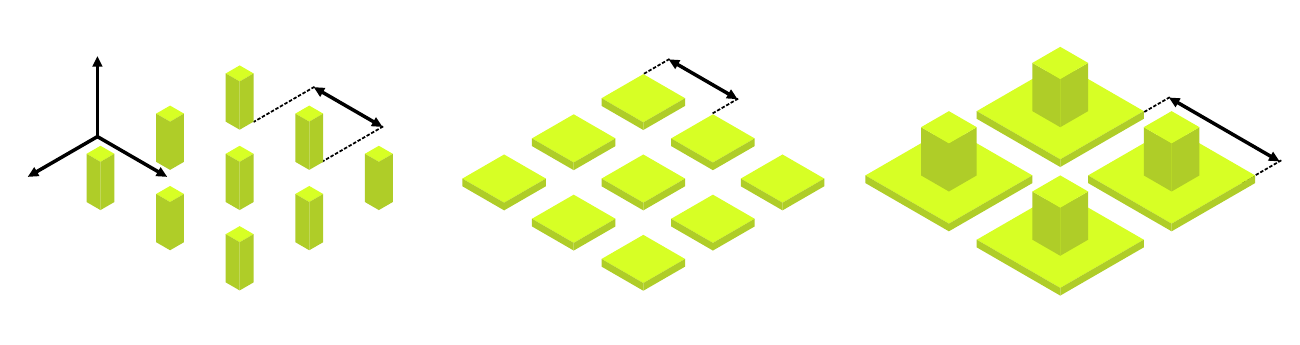}
        \put(19,1){\htext{(a) $\chia{ee}{xx}$, $\chia{mm}{yy}$, and  $\chia{ee}{zz}$}}
        \put(50,1){\htext{(b) $\chia{ee}{xx}$ and $\chia{mm}{yy}$}}
        \put(82,1){\htext{(c) $\chia{ee}{xx}$, $\chia{mm}{yy}$, $\chia{em}{xy}$, $S_\text{me}^{yzzx}$, ...}}
        \put(13.2,12.6){\htext{\small{}$y$}}
        \put(1.4,12){\htext{\small{}$x$}}
        \put(7.5,23){\htext{\small{}$z$}}
        \put(30,20){\htext{\small{}$p\ll\lambda_0$}}
        \put(58,21.5){\htext{\small{}$p\ll\lambda_0$}}
        \put(96.5,18.5){\htext{\small{}$p<\lambda_0$}}
    \end{overpic}
    \caption{Possible unit cells that have given spatial symmetries: (a) and (b) have all the structural symmetries as in Fig. \ref{fig:chi-simplification}c but are deeply subwavelength such that quadrupolar responses are negligible. Since the height of the particles in (b) is negligible, the normal response $\chia{ee}{zz}$ is also negligible in this case. In (c), $\sigma_z$ symmetry is broken, corresponding to Fig. \ref{fig:chi-simplification}d, and is furthermore only slightly subwavelength, meaning that quadrupolar responses are possible.}
    \label{fig:possible-unit-cells}
\end{figure}
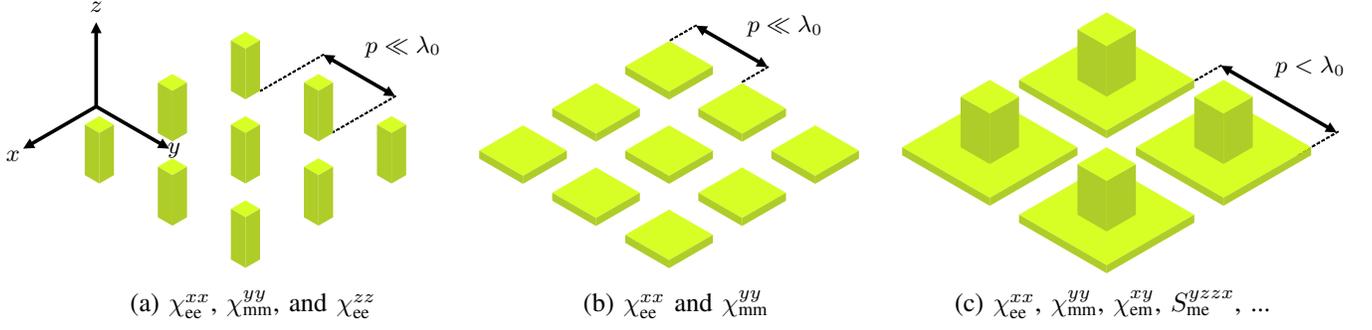

For example, consider the metasurface in Fig.~\ref{fig:possible-unit-cells}a, with all possible symmetries: reflection ($\sigma_x$, $\sigma_y$, $\sigma_z$) and rotation ($C_{4,z}$). These symmetries can be used to write invariance conditions on the susceptibility tensors in \eqref{eq:spatial-dispersion} which eliminate 
incongruous components. The invariance relations are given in \cite{achouriSpatialSymmetriesMultipolar2022} along with an algorithm to easily apply them. Following this algorithm, the hypersusceptiblity matrix reduces to the 9 terms in Fig.~\ref{fig:chi-simplification}c. Furthermore, if the unit cell is deeply subwavelength ($p\ll\lambda$\footnote{$\lambda$ is the wavelength in the background, or the shorter of the two wavelengths in the top or bottom media.}), then higher-order susceptibilities will be negligible such that the surface can be described using only $\chia{ee}{xx}$, $\chia{mm}{yy}$, and $\chia{ee}{zz}$. Also, if the metasurface is very thin, as in \ref{fig:possible-unit-cells}b, then $\chia{ee}{zz}$ may be negligible, such that only $\chia{ee}{xx}$ and $\chia{mm}{yy}$ are necessary.

However, optical meta-atoms are generally large, such that the dipolar model is inappropriate~\cite{achouriMultipolarModelingSpatially2022}. Then, quadrupolar susceptibilities are necessary, and these provide additional degrees of freedom for specifying wave transformations. To provide even more additional degrees of freedom, spatial symmetries can be broken. For example, consider breaking $\sigma_z$ symmetry, as is the case for the meta-atom in Fig.~\ref{fig:possible-unit-cells}c. By following the algorithm in \cite{achouriSpatialSymmetriesMultipolar2022}, one arrives at the matrix with 14 terms in Fig.~\ref{fig:chi-simplification}d. This matrix allows for bianisotropy (e.g. $\chi_\text{em}^{xy}$), and will be used later to demonstrate the utility of the additional degrees of freedom for wave transformations.

\section{Synthesis and Scattering Analysis}
Given (\ref{eq:GSTCs}a-b) and \eqref{eq:consitutive-nodz}, it is possible to calculate the fields that will be scattered from a metasurface. In this section, we will consider how plane waves are scattered for a metasurface and how to engineer the susceptibilities to control the angular scattering behaviour.

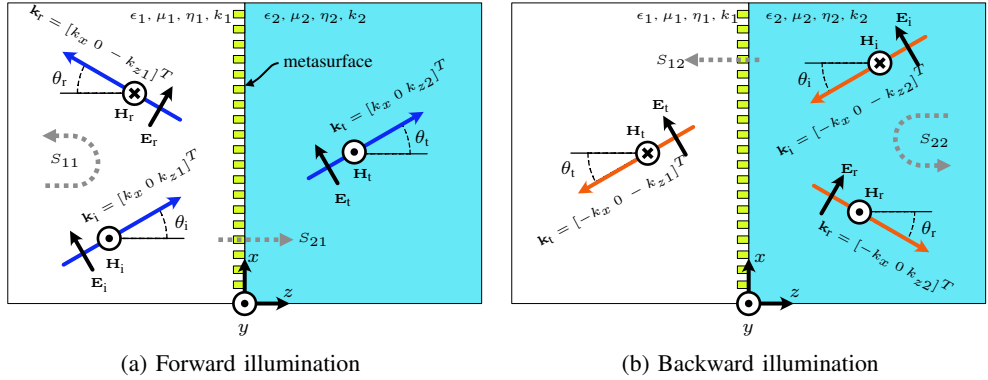
\begin{figure}
    \centering
    \begin{overpic}[grid=false,trim={0cm 0cm 0cm 0cm},clip,tics=2]{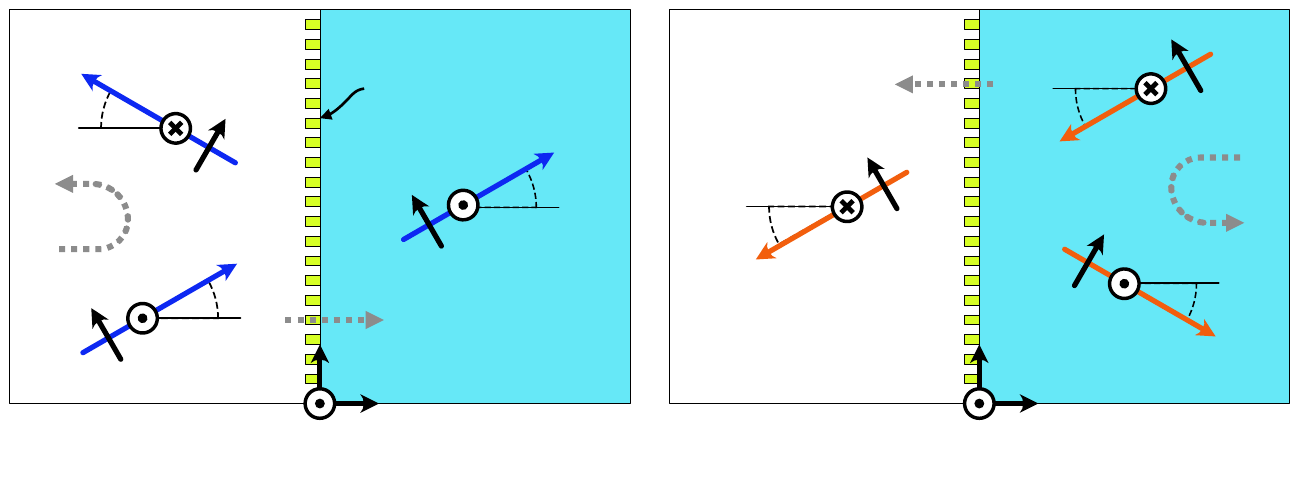}
        \put(24.2,0){\htext{\small{}(a) Forward illumination}}
        \put(75.5,0){\htext{\small{}(b) Backward illumination}}
        \put(25.5,11){\htext{\scriptsize{}$x$}}
        \put(24.5,4){\htext{\scriptsize{}$y$}}
        \put(29,7.5){\htext{\scriptsize{}$z$}}
        \put(76.5,11){\htext{\scriptsize{}$x$}}
        \put(75.3,4){\htext{\scriptsize{}$y$}}
        \put(80,7.5){\htext{\scriptsize{}$z$}}
        \put(18.3,14.5){\htext{\scriptsize{}$\theta_\text{i}$}}
        \put(6,29){\htext{\scriptsize{}$\theta_\text{r}$}}
        \put(42.5,23){\htext{\scriptsize{}$\theta_\text{t}$}}
        \put(81,28.9){\htext{\scriptsize{}$\theta_\text{i}$}}
        \put(93.5,13.8){\htext{\scriptsize{}$\theta_\text{r}$}}
        \put(57.2,19.9){\htext{\scriptsize{}$\theta_\text{t}$}}
        \put(14,18){\makebox[0pt]{\rotatebox[origin=c]{28}{\Centerstack{\tiny{} $\mathbf{k}_\text{i}=\left[k_x \: 0 \: k_{z1}\right]^T$ }}}}
        \put(10,32.5){\makebox[0pt]{\rotatebox[origin=c]{-28}{\Centerstack{\tiny{} $\mathbf{k}_\text{r}=\left[k_x \: 0 \: -k_{z1}\right]^T$ }}}}
        \put(39,27){\makebox[0pt]{\rotatebox[origin=c]{28}{\Centerstack{\tiny{} $\mathbf{k}_\text{t}=\left[k_x \: 0 \: k_{z2}\right]^T$ }}}}
        \put(85.6,25.6){\makebox[0pt]{\rotatebox[origin=c]{28}{\Centerstack{\tiny{} $\mathbf{k}_\text{i}=\left[-k_x \: 0 \: -k_{z2}\right]^T$ }}}}
        \put(89,11){\makebox[0pt]{\rotatebox[origin=c]{-28}{\Centerstack{\tiny{} $\mathbf{k}_\text{r}=\left[-k_x \: 0 \: k_{z2}\right]^T$ }}}}
        \put(61.5,16.5){\makebox[0pt]{\rotatebox[origin=c]{28}{\Centerstack{\tiny{} $\mathbf{k}_\text{t}=\left[-k_x \: 0 \: -{k_{z1}}\right]^T$ }}}}
        \put(11.6,10.4){\htext{\tiny{}$\Hv_\text{i}$}}
        \put(12.5,25.5){\htext{\tiny{}$\Hv_\text{r}$}}
        \put(36.5,19.5){\htext{\tiny{}$\Hv_\text{t}$}}
        \put(10,8.5){\htext{\tiny{}$\Ev_\text{i}$}}
        \put(15,23){\htext{\tiny{}$\Ev_\text{r}$}}
        \put(34.5,17){\htext{\tiny{}$\Ev_\text{t}$}}
        \put(87.5,32.9){\htext{\tiny{}$\Hv_\text{i}$}}
        \put(88,17.6){\htext{\tiny{}$\Hv_\text{r}$}}
        \put(64,23.8){\htext{\tiny{}$\Hv_\text{t}$}}
        \put(91,35.2){\htext{\tiny{}$\Ev_\text{i}$}}
        \put(85.5,20){\htext{\tiny{}$\Ev_\text{r}$}}
        \put(66.5,26.4){\htext{\tiny{}$\Ev_\text{t}$}}
        \put(13,35.5){\tiny{}$\epsilon_1$, $\mu_1$, $\eta_1$, $k_1$}
        \put(26.5,35.5){\tiny{}$\epsilon_2$, $\mu_2$, $\eta_2$, $k_2$}
        \put(64,35.5){\tiny{}$\epsilon_1$, $\mu_1$, $\eta_1$, $k_1$}
        \put(77,35.5){\tiny{}$\epsilon_2$, $\mu_2$, $\eta_2$, $k_2$}
        \put(5,21){\tiny{}$S_{11}$}
        \put(30,13){\tiny{}$S_{21}$}
        \put(92.5,23){\tiny{}$S_{22}$}
        \put(66,31){\tiny{}$S_{12}$}
        \put(28.5,30.7){\scriptsize{}metasurface}
    \end{overpic}
    \caption{A depiction of the TM-polarized plane waves incident along the $xz$ plane onto a metasurface placed at an interface between two different media at $z=0$. The situations for both forward and backward illumination are depicted, with the incident field approaching from the first and second media, respectively.}
    \label{fig:forwards-backwards}
\end{figure}

We will consider TM-polarized plane waves propgagting in the $xz$ plane. With reference Fig.~\ref{fig:forwards-backwards}, we express the fields as
\begin{subequations}
    \begin{align}
        \Hv_\text{a} &= S_\text{a} H_0\mathbf{\hat{y}}e^{-j\mathbf{k}_\text{a}\cdot\mathbf{r}}\\
        \Ev_\text{a} &= \frac{\eta_\text{a}}{k_\text{a}}\Hv_\text{a}\times\mathbf{k}_\text{a}
    \end{align}
\end{subequations}
with $\text{a}=\text{i}$ for the incident field (fields are normalized with $S_\text{i}=1$), $\text{a}=\text{r}$ for the reflected field ($S_\text{r}=-S_{11}$, the reflection coefficient), and $\text{a}=\text{t}$ for the transmitted field ($S_\text{t}=S_{21}$, the transmission coefficient). For backwards illumination, one replaces $1\Longleftrightarrow 2$ along with $k_x\rightarrow -k_x$, and $k_{z,\{1,2\}}\rightarrow -k_{z,\{2,1\}}$. This is shown in Fig.~\ref{fig:forwards-backwards}.

Now, these fields can be substituted into the GSTCs (\ref{eq:GSTCs}a-b) and the constitutive relations \eqref{eq:consitutive-nodz}. In the case of forward illumination, this provides equations to solve for the two unknowns $S_{11}$ and $S_{21}$, and for backwards illumination one can solve for $S_{22}$ and $S_{21}$. However, the expressions for these S-parameters are very unwieldy, and so we will limit the analysis to some of the terms from Fig.~\ref{fig:possible-unit-cells}. Considering the dipolar susceptibilities $\chia{ee}{xx}$, $\chia{mm}{yy}$, $\chia{ee}{zz}$, then
\begin{subequations}
\label{eq_SXeXm}
    \begin{gather}
        S_{\{11,22\}}(k_x) = -\frac{k_{z,\{1,2\}}\alpha + k_{z,\{2,1\}}\left( 4jk_{z,\{1,2\}}\chia{ee}{xx} +n_{\{1,2\}}^2 \alpha \right) + 4jn_{\{1,2\}}^2\left( k_x^2\chia{ee}{zz} + k_0^2\chia{mm}{yy} \right)
        }{k_{z,\{1,2\}}\alpha + k_{z,\{2,1\}}\left( 4jk_{z,\{1,2\}}\chia{ee}{xx} +n_{\{1,2\}}^2 \alpha \right) + 4jn_{\{1,2\}}^2\left( k_x^2\chia{ee}{zz} - k_0^2\chia{mm}{yy} \right)}\\
        S_{\{21,12\}}(k_x) = \frac{2k_{z,\{1,2\}}n_1n_2\alpha
        }{k_{z,\{1,2\}}\alpha + k_{z,\{2,1\}}\left( 4jk_{z,\{1,2\}}\chia{ee}{xx} +n_{\{1,2\}}^2 \alpha \right) + 4jn_{\{1,2\}}^2\left( k_x^2\chia{ee}{zz} - k_0^2\chia{mm}{yy} \right)}\\
        \alpha = k_x^2\chia{ee}{xx}\chia{ee}{zz} + k_0^2\chia{ee}{xx}\chia{mm}{yy} - 4 \,,
    \end{gather}
\end{subequations}
with the first subscripts selected for forward illumination ($S_{11}$ and $S_{21}$) and the second for backward illumination ($S_{22}$ and $S_{12}$).

Meanwhile, for use later, we will also derive expressions for the scattering with $\chia{em}{xy}$, $S_\text{me}^{yzzx}$. We find
\begin{subequations}
    \begin{align}
        S_{\{11,22\}}(k_x) = \frac{\mp k_{z,1}\eta_2^2\beta^{\pm}\pm k_{z,2}\eta_1^2\beta^{\mp}}
        {k_{z,1}\eta_2\beta^{\pm} + k_{z,2}\eta_1\beta^{\mp}}
    \end{align}%
    \begin{align}
        S_{\{21,12\}}(k_x) = \frac{2n_1n_2k_{z,\{1,2\}}\left[\splitfrac{
            4k_x^4(S_\text{me}^{yzzx})^2+k_0^4(S_\text{me}^{yzzx}+4\chia{em}{xy})^2 }{-4k_0^2\left(k_x^2S_\text{me}^{yzzx}(S_\text{me}^{yzzx}+4\chia{em}{xy})-16\right)}\right]
        }
        {k_{z,1}\eta_2\beta^{\pm} + k_{z,2}\eta_1\beta^{\mp}}
    \end{align}%
    \begin{align}
        \beta^{\pm}=\left[8jk_0\pm 2k_x^2S_\text{me}^{yzzx}\mp k_0^2(S_\text{me}^{yzzx}+4\chia{em}{xy})\right]^2 \,,
    \end{align}
    \label{eq:sparams}
\end{subequations}
where the top sign is selected for forward illumination ($S_{11}$, $S_{21}$) and the bottom sign is selected for backward illumination ($S_{22}$, $S_{12}$).

Given these expressions, one can solve for the susceptibilities required to control the angular scattering behaviour. For example, to suppress reflection at some angle $k_x^{'}$, one needs to solve $|S_{11}(k_x^{'})|=0$.

\section{Illustration: Tuning Brewster's Angle}
\label{sec:brewster}
\begin{figure}
    \centering
    \begin{overpic}[grid=false,trim={0cm 0cm 0cm 0cm},clip,tics=5]{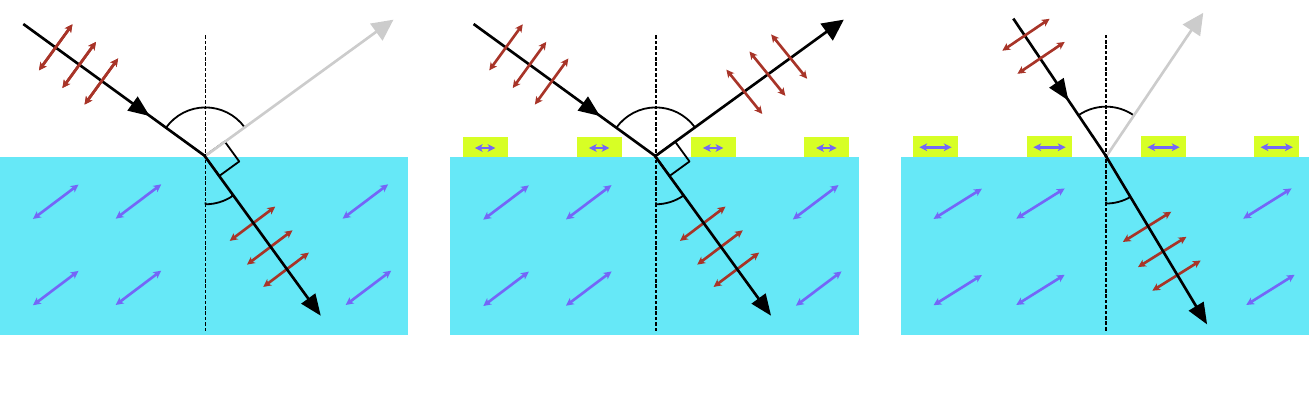}
        \put(16,0){\htext{\small{}(a)}}
        \put(50,0){\htext{\small{}(b)}}
        \put(84,0){\htext{\small{}(c)}}
        
        \put(13.5,22.5){\htext{\scriptsize{}$\theta_\text{i,1}$}}
        \put(17.5,22.5){\htext{\scriptsize{}$\theta_\text{r,1}$}}
        \put(17,13){\htext{\scriptsize{}$\theta_\text{t,1}$}}
        \put(8,11){\htext{\scriptsize{}$\Pv_\text{bulk}$ ($\parallel \Ev_\text{r}$)}}
        \put(7,28){\htext{\scriptsize{}$\Ev_\text{i}$}}
        \put(17.5,9){\htext{\scriptsize{}$\Ev_\text{t}$}}
        \put(24,27){\htext{\scriptsize{}$\Ev_\text{r}=0$}}
        \put(48,22.5){\htext{\scriptsize{}$\theta_\text{i,1}$}}
        \put(52,22.5){\htext{\scriptsize{}$\theta_\text{r,1}$}}
        \put(51.5,13){\htext{\scriptsize{}$\theta_\text{t,1}$}}
        \put(42,11){\htext{\scriptsize{}$\Pv_\text{bulk}$}}
        \put(37,20.5){\htext{\scriptsize{}$\Pv_\text{MS}$}}
        \put(42,28){\htext{\scriptsize{}$\Ev_\text{i,1}$}}
        \put(52,9){\htext{\scriptsize{}$\Ev_\text{t,1}$}}
        \put(56.5,28){\htext{\scriptsize{}$\Ev_\text{r,1}\neq 0$}}
        \put(83,22.5){\htext{\scriptsize{}$\theta_\text{i,2}$}}
        \put(86,22.5){\htext{\scriptsize{}$\theta_\text{r,2}$}}
        \put(85.5,13){\htext{\scriptsize{}$\theta_\text{t,2}$}}
        \put(75.7,24.5){\htext{\scriptsize{}$\Ev_\text{i,2}$}}
        \put(86.5,7){\htext{\scriptsize{}$\Ev_\text{t,2}$}}
        \put(93,24.5){\htext{\scriptsize{}$\Ev_\text{r,2}=0$}}
        \put(76,11){\htext{\scriptsize{}$\Pv_\text{bulk}$}}
        \put(71.5,20.5){\htext{\scriptsize{}$\Pv_\text{MS}$}}
    \end{overpic}
    \caption{For ordinary dielectric media, the Brewster angle occurs when the reflected and refracted rays are orthogonal and thus prohibit any reflection, as in (a). If a metasurface is introduced, the additional polarization at the surface contributes to the reflected field and thus they can be non-zero, as in (b). However, the Brewster angle now can occur at other angle(s), when the net scattering due to the substrate and metasurface interfere destructively, as in (c).}
    \label{fig:brewster-illustration}
\end{figure}
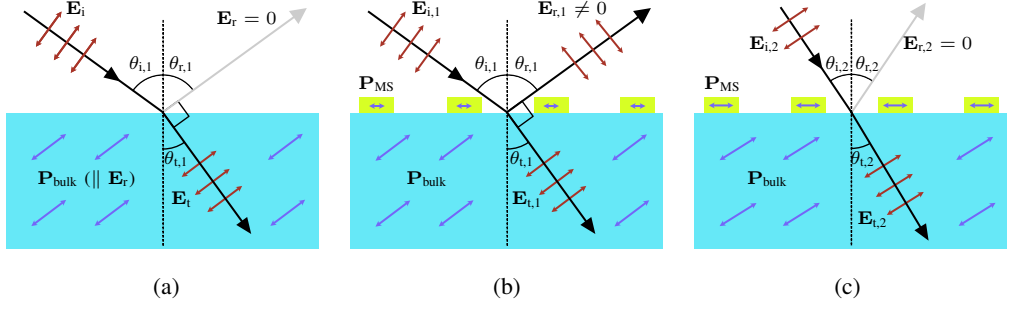

The Brewster angle is defined as the angle of incidence of a plane wave at a dielectric interface where reflection is eliminated; i.e., there is complete transmission. For ordinary (non-magnetic) materials, it only occurs for TM polarization, and can be understood intuitively from Fig.~\ref{fig:brewster-illustration}a. When the angle of incidence is $\theta_\text{i,1}$ and the angle of refraction is $\theta_\text{t,1}$, the wave vectors of the refracted and reflected fields are orthogonal. Consequently, the bulk electric polarization in the substrate is orthogonal to the electric field of the reflected field and it is impossible for the bulk polarization to produce a reflected field. This Brewster angle $\theta_\text{i}=\theta_\text{B}$ occurs at \cite{balanisAdvancedEngineeringElectromagnetics2012}
\begin{gather}
    \theta_\text{B}=\tan^{-1}
    \sqrt{\frac{\epsilon_2}{\epsilon_1}} \,,\label{eq:brewster-ordinary}
\end{gather}
which corresponds to $k_{x}=k_{\text{B}}=k_1\sin\theta_\text{B}$.

Using a metasurface, it is possible to tune $\theta_\text{B}$ \cite{lavigneGeneralizedBrewsterEffect2021}. In Fig.~\ref{fig:brewster-illustration}b, a metasurface has been added that has a surface polarization which has a non-zero projection onto the reflected electric field at $\theta_\text{i,1}$ and thus there reflection is no longer suppressed. However, now there can be another angle $\theta_\text{i,2}$ where the superposition of the scattered fields from the bulk and the metasurface result in a suppression of the scattered fields, as in Fig.~\ref{fig:brewster-illustration}c. By adjusting the metasurface, this angle can be tuned. In fact, we will show that it is possible to have multiple such Brewster angles. There is one caveat: metasurfaces are generally resonant and thus limited in bandwidth unlike the broadband dielectric Brewster effect given by \eqref{eq:brewster-ordinary}. 

In this section, we will illustrate the use of the additional susceptibility components from Fig.~\ref{fig:chi-simplification}d to achieve this generalized Brewster effect.

\subsection{Generalized Brewster Effect}

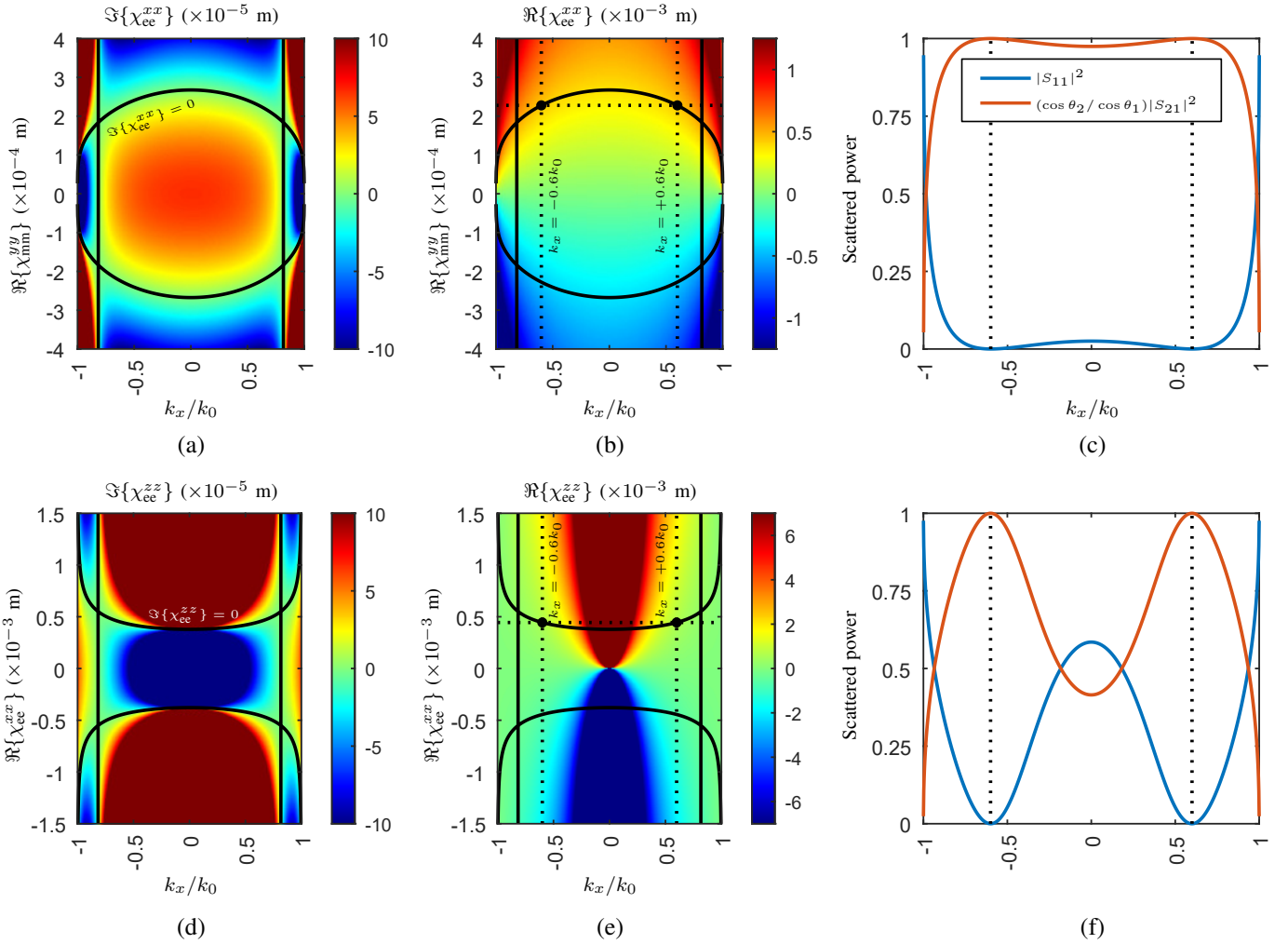
\begin{figure}
    \centering
    \begin{overpic}[width=\textwidth,grid=false,trim={0cm 0cm 0cm 0cm},clip]{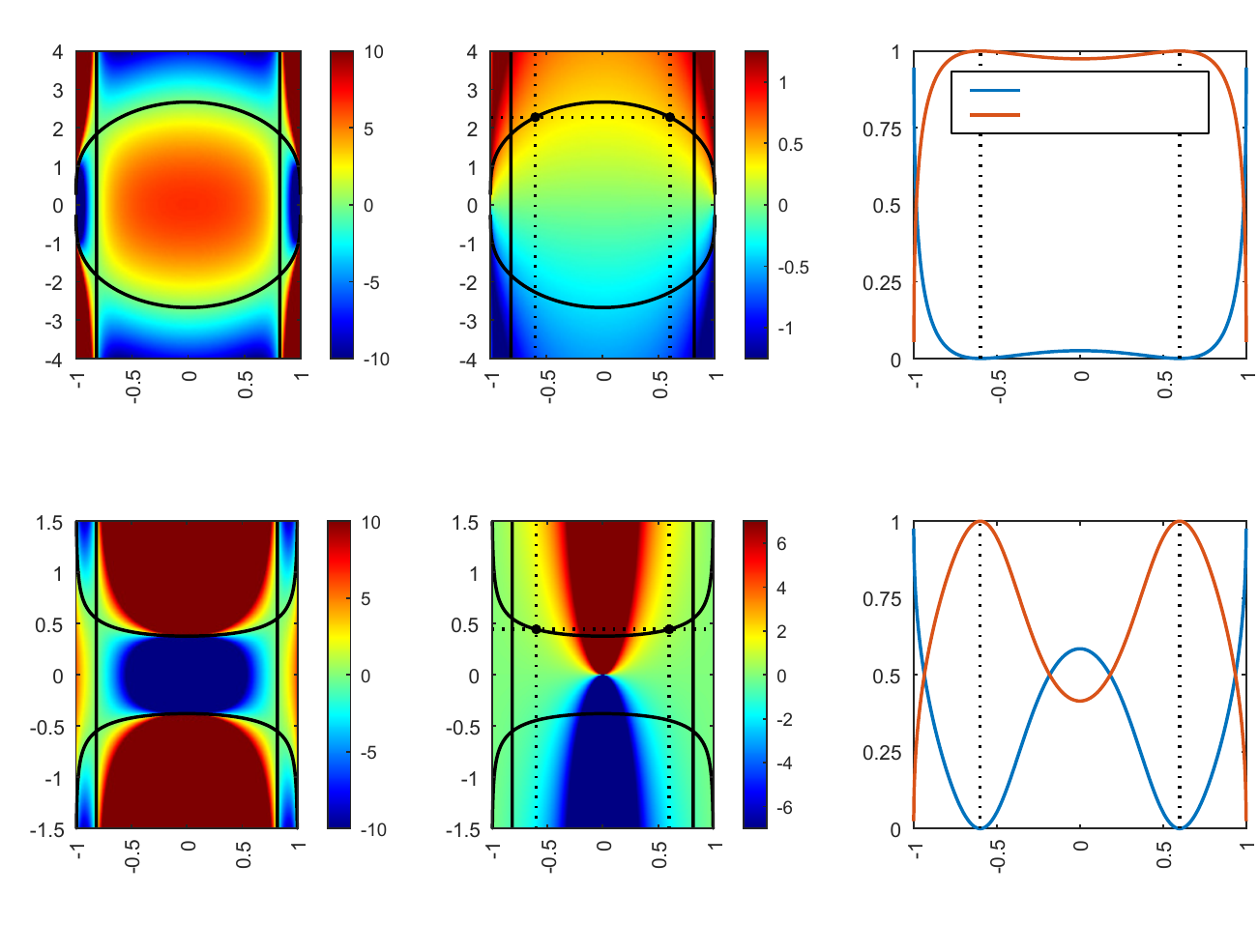}
        \put(15,1){\htext{(d)}}
        \put(48,1){\htext{(e)}}
        \put(86,1){\htext{(f)}}
        \put(15,39){\htext{(a)}}
        \put(48,39){\htext{(b)}}
        \put(86,39){\htext{(c)}}
        \put(15,4.5){\htext{\footnotesize{}$k_x/k_0$}}
        \put(1,21){\vtext{\footnotesize{}$\Re\{\chia{ee}{xx}\}$ ($\times 10^{-3}$ m)}}
        \put(15,35.5){\htext{\footnotesize{}$\Im\{\chia{ee}{zz}\}$ ($\times 10^{-5}$ m)}}
        \put(48,4.5){\htext{\footnotesize{}$k_x/k_0$}}
        \put(34,21){\vtext{\footnotesize{}$\Re\{\chia{ee}{xx}\}$ ($\times 10^{-3}$ m)}}
        \put(48,35.5){\htext{\footnotesize{}$\Re\{\chia{ee}{zz}\}$ ($\times 10^{-3}$ m)}}
        \put(15,42){\htext{\footnotesize{}$k_x/k_0$}}
        \put(1.5,58){\vtext{\footnotesize{}$\Re\{\chia{mm}{yy}\}$ ($\times 10^{-4}$ m)}}
        \put(15,73){\htext{\footnotesize{}$\Im\{\chia{ee}{xx}\}$ ($\times 10^{-5}$ m)}}
        \put(48,42){\htext{\footnotesize{}$k_x/k_0$}}
        \put(34.5,58){\vtext{\footnotesize{}$\Re\{\chia{mm}{yy}\}$ ($\times 10^{-4}$ m)}}
        \put(48,73){\htext{\footnotesize{}$\Re\{\chia{ee}{xx}\}$ ($\times 10^{-3}$ m)}}
        %
        %
        \put(86,42){\htext{\footnotesize{}$k_x/k_0$}}
        \put(81.5,68.2){\tiny{}$|S_{11}|^2$}
        \put(81.5,66.1){\tiny{}$(\cos\theta_2/\cos\theta_1)|S_{21}|^2$}
        \put(11.6,65){\makebox[0pt]{\rotatebox[origin=c]{20}{\Centerstack{\tiny{} $\Im\{\chia{ee}{xx}\}=0$ }}}}
        \put(15,26){\makebox[0pt]{\rotatebox[origin=c]{0}{\Centerstack{\tiny{} \color{white}$\Im\{\chia{ee}{zz}\}=0$ }}}}
        \put(43.4,58){\vtext{\color{black}\tiny{}$k_x=-0.6k_0$}}
        \put(51.8,58){\vtext{\color{black}\tiny{}$k_x=+0.6k_0$}}
        \put(43.4,29.3){\vtext{\color{black}\tiny{}$k_x=-0.6k_0$}}
        \put(51.8,29.3){\vtext{\color{black}\tiny{}$k_x=+0.6k_0$}}
        \put(66.9,21){\vtext{\footnotesize{}Scattered power}}
        \put(66.9,58.5){\vtext{\footnotesize{}Scattered power}}
    \end{overpic}
    \caption{Brewster angle control using dipolar susceptibilites, considering an interface with two media having $\epsilon_1=1$ and $\epsilon_2=2$ and at $\lambda_0=c_0/(\SI{300}{THz})$. (a-b) A depiction of the real of imaginary part of $\chia{ee}{xx}$ which satisfy \eqref{eq:brewster-condition-dip-tang} given a desired $k_{\text{B}}$ and $\chia{mm}{yy}\in\mathbb{R}$. (c) Reflected and transmitted power in the case of $\chia{ee}{xx}=\SI{4.44e-4}{m}$ and $\chia{mm}{yy}=\SI{2.28e-4}{m}$, where $k_{x,\text{B}}=0.6k_0$. (a-c) The same plots, but considering $\chia{ee}{xx}$ and $\chia{ee}{zz}$, with $\chia{ee}{xx}=\SI{4.44e-4}{m}$ and $\chia{ee}{zz}=\SI{6.34e-4}{m}$ in (f).}
    \label{fig:brewster-dipolar}
\end{figure}

Consider the very simple case of a Huygen's metasurface with $\chia{ee}{xx}$ and $\chia{mm}{yy}$. Using these two terms, it is possible to achieve a single wave transformation \cite{achouriGeneralMetasurfaceSynthesis2015}, which we would desire to be the suppression of reflection at a given angle of incidence. This corresponds to impedance matching the two media at a given angle. We will ensure that the metasurface is lossless, which will ensure that all power is transmitted. Setting $|S_{21}|=0$ in (\ref{eq:sparams}) and solving for $\chia{ee}{xx}$, we find
\begin{align}
    \chia{ee}{xx} = \frac{4\eta_0(k_2k_{z1}\eta_1-k_1k_{z2}\eta_2-jk_0k_1k_2\chia{mm}{yy})}{k_0[k_0\chia{mm}{yy}\eta_0(k_2k_{z1}\eta_1-k_1k_{z2}\eta_2)-4j\eta_1\eta_2k_{z1}k_{z2}]} \,,
    \label{eq:brewster-condition-dip-tang}
\end{align}
which shows that $\chia{ee}{xx}$ must in general be complex, implying a metasurface with loss and gain, as was noted in \cite{lavigneGeneralizedBrewsterEffect2021}. However, we will show that it is possible to avoid the need for loss or gain. In order for the metasurface to be lossless, $\chia{ee}{xx}$ and $\chia{mm}{yy}$ must both be purely real. To this end, we will set $\chia{mm}{yy}$ to be a purely real number and then find $\chia{ee}{xx}$ from \eqref{eq:brewster-condition-dip-tang} which are also purely real.

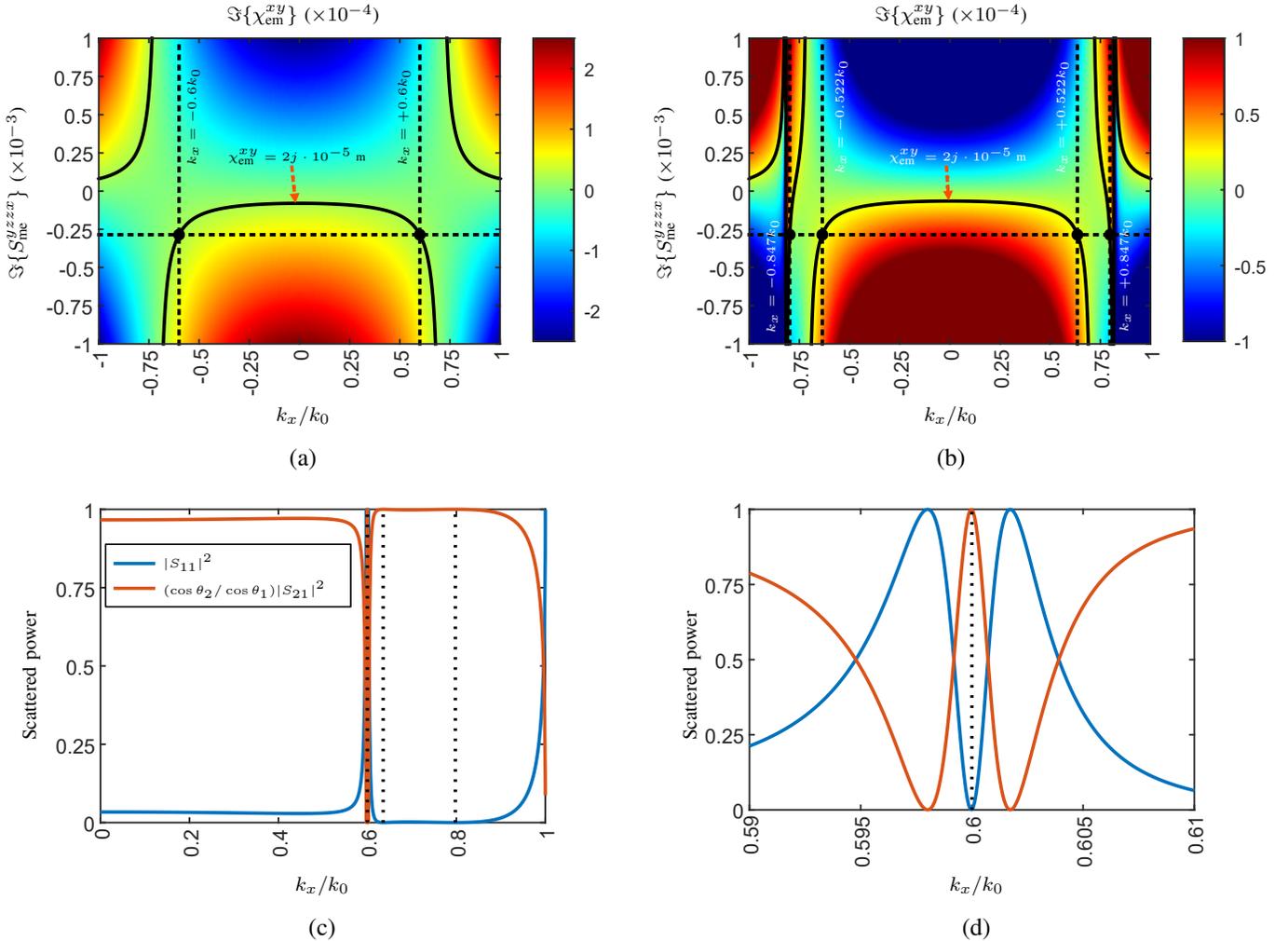
\begin{figure}[b!]
    \centering
    \begin{overpic}[width=\textwidth,grid=false,trim={0cm 0cm 0cm 0cm},clip,tics=5]{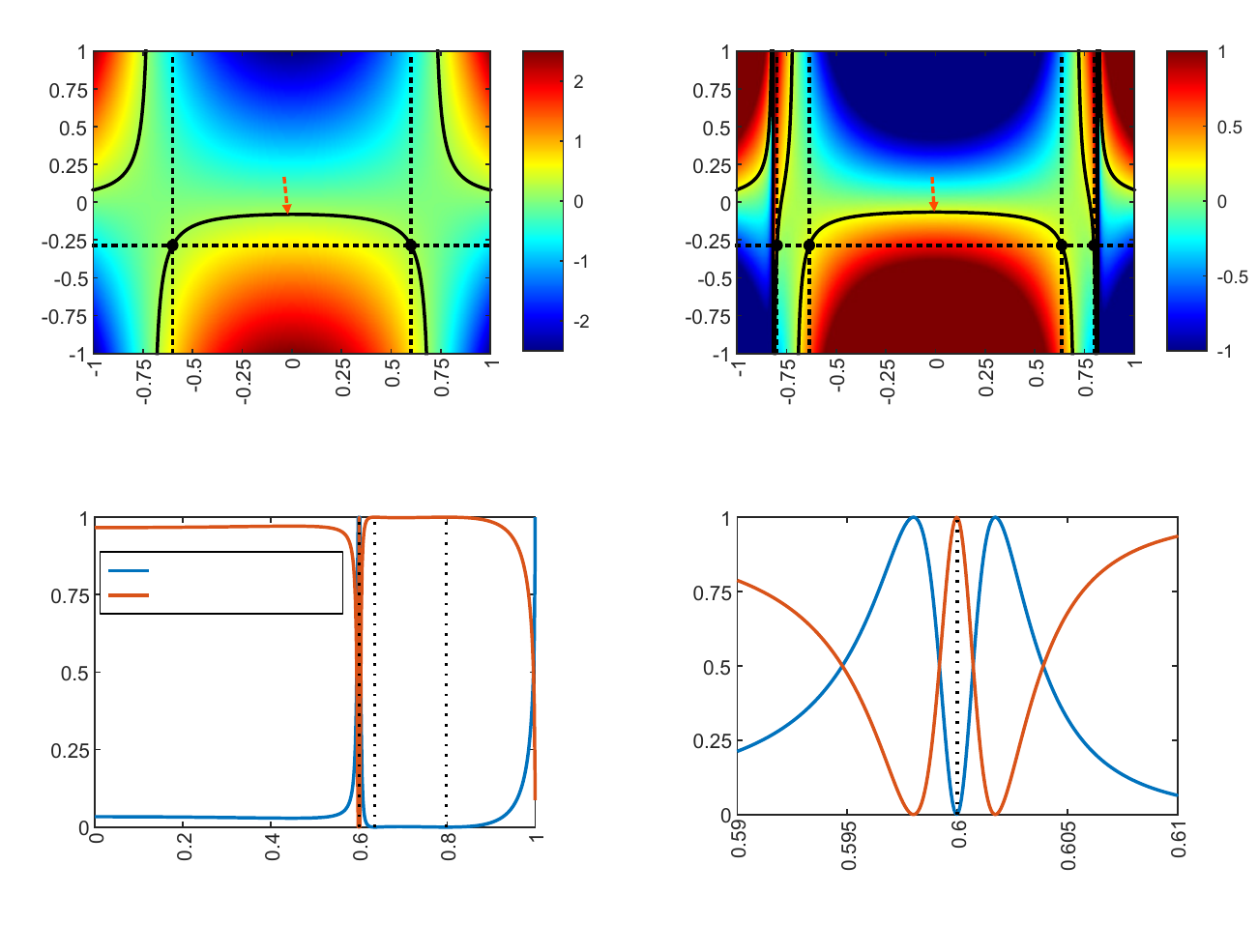}
        \put(25,1){\htext{(c)}}
        \put(76.5,1){\htext{(d)}}
        \put(23.5,38){\htext{(a)}}
        \put(74.5,38){\htext{(b)}}
        \put(25,4.5){\htext{\footnotesize{}$k_x/k_0$}}
        %
        \put(76.5,4.5){\htext{\footnotesize{}$k_x/k_0$}}
        \put(23.5,41.5){\htext{\footnotesize{}$k_x/k_0$}}
        \put(1.0,59){\vtext{\footnotesize{}$\Im\{S_\text{me}^{yzzx}\}$ ($\times 10^{-3}$)}}
        \put(23.5,73){\htext{\footnotesize{}$\Im\{\chia{em}{xy}\}$ ($\times 10^{-4}$)}}
        \put(74.5,41.5){\htext{\footnotesize{}$k_x/k_0$}}
        \put(52,59){\vtext{\footnotesize{}$\Im\{S_\text{me}^{yzzx}\}$ ($\times 10^{-3}$)}}
        \put(74.5,73){\htext{\footnotesize{}$\Im\{\chia{em}{xy}\}$ ($\times 10^{-4}$)}}
        \put(12.5,30){\tiny{}$|S_{11}|^2$}
        \put(12.5,27.9){\tiny{}$(\cos\theta_2/\cos\theta_1)|S_{21}|^2$}
        %
        \put(14.6,65){\vtext{\tiny{}$k_x=-0.6k_0$}}
        \put(31.2,65){\vtext{\tiny{}$k_x=+0.6k_0$}}
        %
        %
        \put(60.2,52.5){\vtext{\color{white}\tiny{}$k_x=-0.847k_0$}}
        \put(65.6,65){\vtext{\color{white}\tiny{}$k_x=-0.522k_0$}}
        \put(83,65){\vtext{\color{white}\tiny{}$k_x=+0.522k_0$}}
        \put(88,52.5){\vtext{\color{white}\tiny{}$k_x=+0.847k_0$}}
        \put(17.6,62){\tiny{}$\chia{em}{xy}=2j\cdot 10^{-5}$ m}
        \put(69.6,62){\color{white}\tiny{}$\chia{em}{xy}=2j\cdot 10^{-5}$ m}
        \put(2,21){\vtext{\footnotesize{}Scattered power}}
        \put(54,21){\vtext{\footnotesize{}Scattered power}}
        %
        %
    \end{overpic}
    \caption{By using quadrupolar susceptibilities, multiple Brewester angles are predicted, with
    \eqref{eq:antibrewster-quadrupolar} having two solutions for $\chia{em}{xy}$ plotted in (a) and (b). Both susceptibilities $\chia{em}{xy}$ and $S_\text{me}^{yzzx}$ are purely imaginary; i.e. lossless, and the black contours correspond to $\chia{em}{xy}=2.00j{\times} 10^{-5}$. Next, selecting $S_\text{me}^{yzzx}=-0.285j{\times} 10^{-3}$ m and $\chia{em}{xy}=2.00j{\times} 10^{-5}$ m, the transmitted and reflected power is plotted in (c) and (d). Note: $\epsilon_1=1$, $\epsilon_2=2$ and $\lambda_0=c_0/(\SI{300}{THz})$.}
    \label{fig:brewster-quad}
\end{figure}

Figure~\ref{fig:brewster-dipolar}a shows $\Im\{\chia{ee}{xx}\}$ plotted as a function of the angle of incidence ($k_x/k_0$) on the $x$-axis and $\chia{mm}{yy}$ on the $y$-axis, for a particular case  where a plane wave is incident from air ($\epsilon_1=1$) onto a substrate ($\epsilon_2=2$).  Consider in particular the black contours along which $\Im\{\chia{ee}{xx}\}=0$ Choosing a point along these contours corresponds to a lossless and gainless metasurface. Next, Fig.~\ref{fig:brewster-dipolar}b shows the same contours superimposed on a plot of $\Re\{\chia{ee}{xx}\}$. We see that the contours cover all incident angles ($-k_0<k_x<k_0$). Thus, one can choose any angle of incidence and the necessary susceptibilities to achieve a Brewster effect at the given angle. For example, to achieve $k_\text{B}=0.6k_0$, $\chia{ee}{xx}=\SI{4.44e-4}{m}$ and $\chia{mm}{yy}=\SI{2.28e-4}{m}$. The corresponding reflected and transmitted power is plotted in Fig.~\ref{fig:brewster-dipolar}c, corroborating a Brewster angle at $k_\text{B}=0.6k_0$.\footnote{For the transmitted power, an angle-dependent factor is used in $(\cos\theta_2/\cos\theta_1)|S_{21}|^2$ to project the Poynting vector to $\zv$ \cite{achouriElectromagneticMetasurfacesTheory2021}.} To practically realize such a metasurface, one could use the cell in Fig.~\ref{fig:possible-unit-cells}b.

Now, we will consider other susceptibility terms to demonstrate how the additional degrees of freedom provide more control over the Brewster effect. Motivated by the scattering caused by $\chia{ee}{zz}$, which is proportional to $\sin^2(\theta_\text{B})\cdot\chia{ee}{zz}$ (unlike $\cos^2(\theta_\text{B})\cdot\chia{ee}{xx}$ and $1\cdot\chia{mm}{yy}$), which can be deduced from~\eqref{eq_SXeXm}, let us consider the combination of $\chia{ee}{zz}$ and $\chia{ee}{xx}$. This could be achieved using a meta-atom like that in Fig.~\ref{fig:possible-unit-cells}a.\footnote{Though $\chia{mm}{yy}$ may in general be present as well, it will typically have a Lorentzian wavelength dependence such that the metasurface can be designed to operate at a frequency where it is negligible \cite{achouriElectromagneticMetasurfacesTheory2021}.}

In this case, setting $|S_{21}|=0$ in (\ref{eq:sparams}) and solving for $\chia{ee}{zz}$, we find
\begin{align}
    \chia{ee}{zz}=
    \frac{4k_0( k_2k_{z1}\eta_0\eta_1 - k_1k_{z2}\eta_0\eta_2 + jk_0k_{z1}k_{z2}\eta_1\eta_2\chia{ee}{xx})}
    {k_x^2\eta_0\left[ 4jk_1k_2\eta_0 +k_0(k_2k_{z1}\eta_1-k_1k_{z2}\eta_2)\chia{ee}{xx} \right]}
    \label{eq:brewster-condition-dip-normal}
\end{align}
for which the real and imaginary parts are plotted in Fig.~\ref{fig:brewster-dipolar}d and e. In this case, designing for $k_\text{B}=0.6k_0$ results in the reflection plotted in Fig.~\ref{fig:brewster-dipolar}f. Compared to the use of $\chia{mm}{yy}$ and $\chia{ee}{xx}$, the use of $\chia{ee}{zz}$ results in a much sharper reflection minimum around the Brewster angle.

Finally, we consider the use of quadrupolar susceptibilities. Using the meta-atom of Fig.~\ref{fig:possible-unit-cells}c, many terms are possible. To demonstrate the versatility of these terms, we will apply two in particular: $S_\text{me}^{yzzx}$ and $\chia{em}{xy}$.\footnote{We assume all other terms but these two are negligible. We leave the design of such a metasurface, which is not a trivial task, as future work.} Once again, setting $|S_{21}|=0$ in (\ref{eq:sparams}) and solving for $\chia{em}{xy}$, one finds
\begin{align}
    \chia{em}{xy} = 
    \frac{\left(2k_x^2-k_0^2\right)S_\text{me}^{yzzx}}{4k_0^2}-
    \frac{\left(\sqrt{k_2k_{z1}\eta_1}\pm\sqrt{k_1k_{z2}\eta_2}\right)^2}{2jk_0\left(k_2k_{z1}\eta_1-k_1k_{z2}\eta_2\right)}
    \label{eq:brewster-quad}
\end{align}
where, noting that the last term is imaginary, then $\chia{em}{xy}$ will be imaginary if $S_\text{me}^{yzzx}$ is imaginary. This corresponds to a lossless metasurface \cite{achouriExtensionLorentzReciprocity2021}, and so we have two solutions. These are plotted in Fig.~\ref{fig:brewster-quad}a and b. To design for $k_{x,\text{B}}=0.6k_0$, we first arbitrarily select $\chia{em}{xy}=2j\cdot 10^{-5}$, which is indicated with black contours. Then, $S_\text{me}^{yzzx}$ is selected using the first solution in Fig.~\ref{fig:brewster-quad}a. However, there are two more angles predicted by the second solution, resulting into three Brewster angles. The scattered power is plotted in Fig.~\ref{fig:brewster-quad}c and d, corroborating the presence of three Brewster angles. Note that the minimum in reflection at $0.6k_0$ is very sharp.

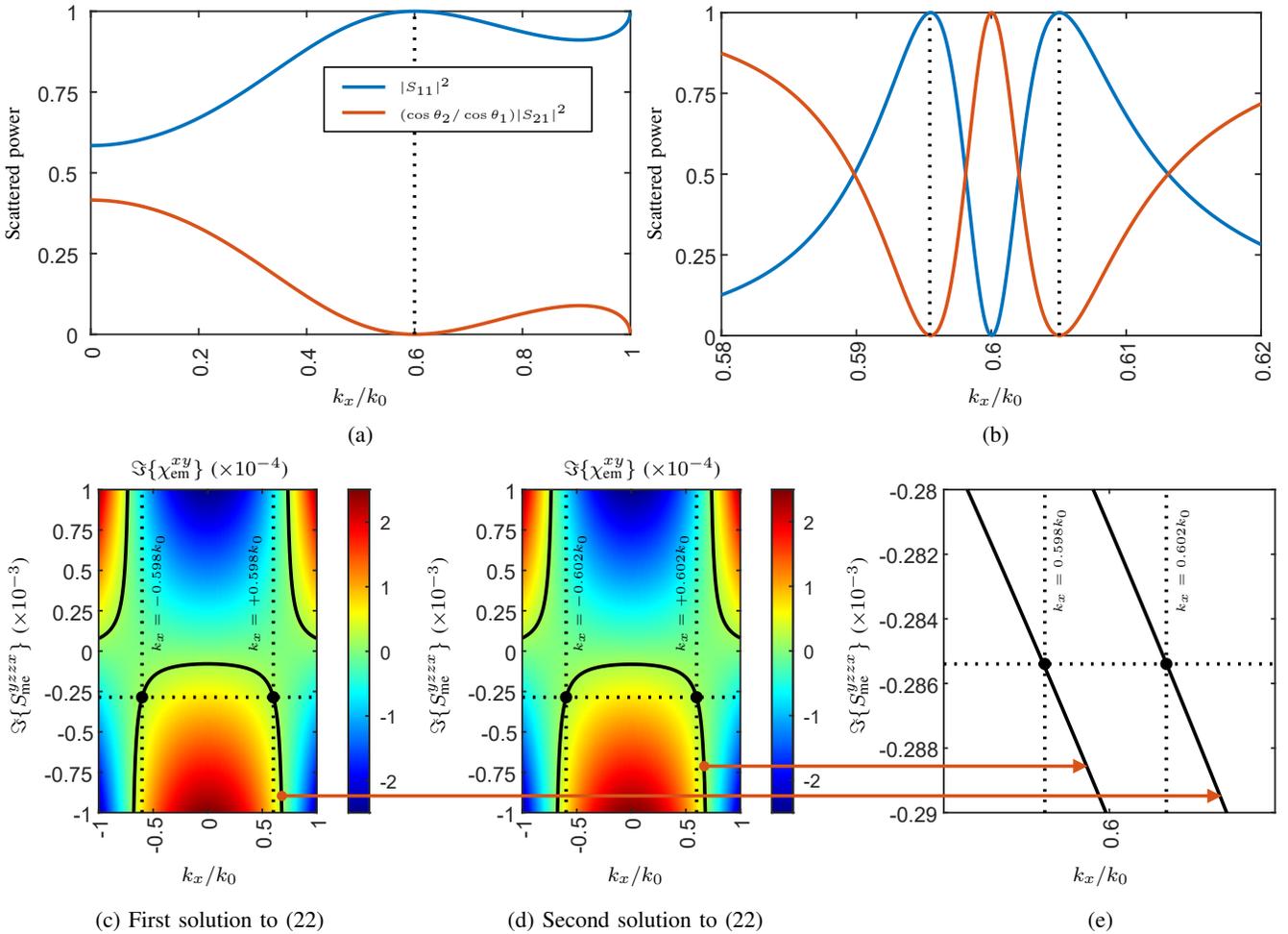
\begin{figure}[b!]
    \centering
    \begin{overpic}[width=\textwidth,grid=false,trim={0cm 0cm 0cm 0cm},clip]{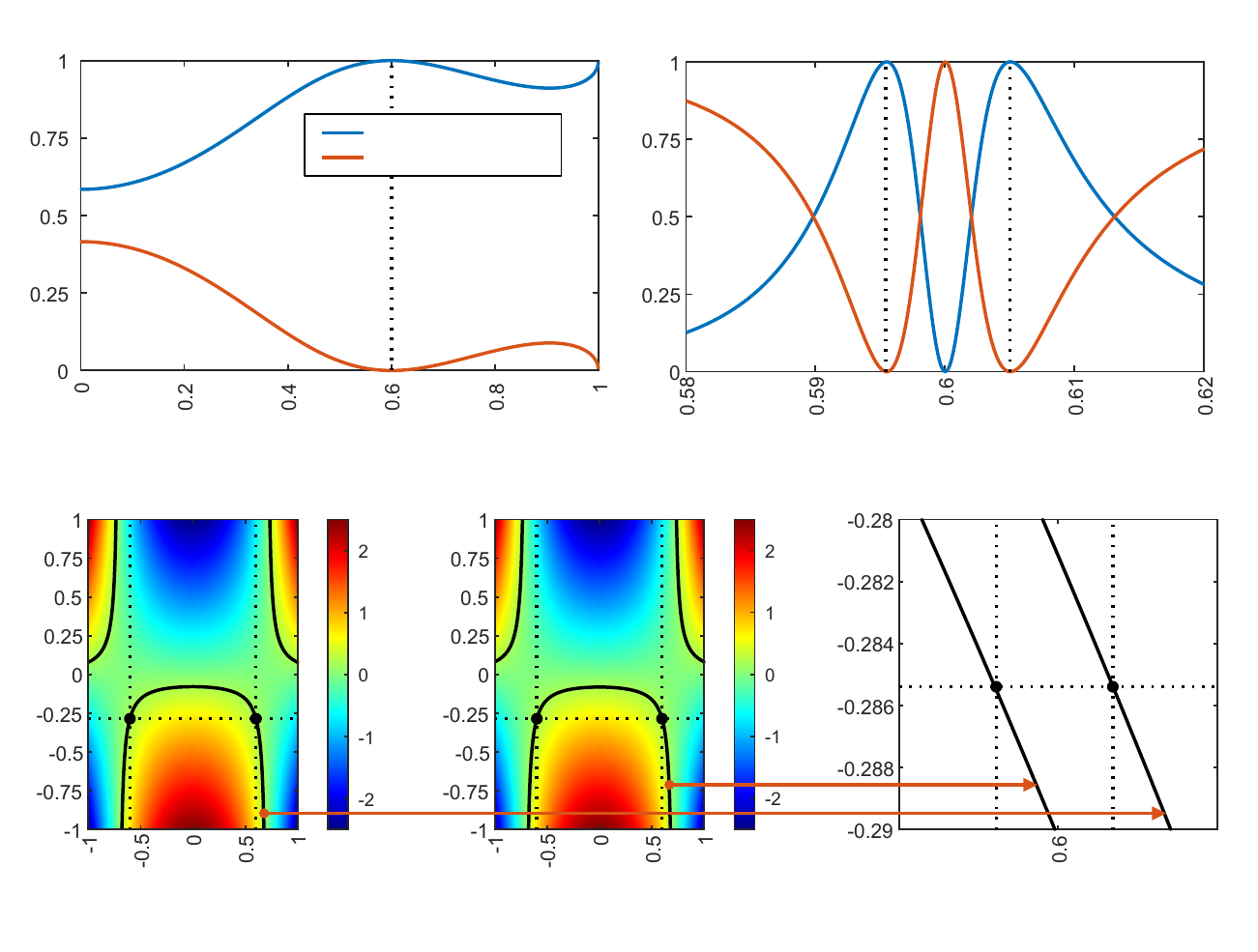}
     \put(15.5,1){\htext{\small{}(c) First solution to \eqref{eq:antibrewster-quadrupolar}}}
     \put(48,1){\small{}\htext{(d) Second solution to \eqref{eq:antibrewster-quadrupolar}}}
     \put(83.5,1){\small{}\htext{(e)}}
     \put(27,38){\small{}\htext{(a)}}
     \put(75.5,38){\small{}\htext{(b)}}
     \put(15.5,4.5){\htext{\footnotesize{}$k_x/k_0$}}
     \put(48,4.5){\htext{\footnotesize{}$k_x/k_0$}}
     \put(83.5,4.5){\htext{\footnotesize{}$k_x/k_0$}}
     \put(27,41){\htext{\footnotesize{}$k_x/k_0$}}
     \put(75.5,41){\htext{\footnotesize{}$k_x/k_0$}}
    \put(1,22){\vtext{\footnotesize{}$\Im\{S_\text{me}^{yzzx}\}$ ($\times 10^{-3}$)}}
    \put(15.5,35.5){\htext{\footnotesize{}$\Im\{\chia{em}{xy}\}$ ($\times 10^{-4}$)}}
    \put(33,22){\vtext{\footnotesize{}$\Im\{S_\text{me}^{yzzx}\}$ ($\times 10^{-3}$)}}
    \put(48.5,35.5){\htext{\footnotesize{}$\Im\{\chia{em}{xy}\}$ ($\times 10^{-4}$)}}
    \put(65,22){\vtext{\footnotesize{}$\Im\{S_\text{me}^{yzzx}\}$ ($\times 10^{-3}$)}}
    \put(30,64.8){\tiny{}$|S_{11}|^2$}
    \put(30,62.7){\tiny{}$(\cos\theta_2/\cos\theta_1)|S_{21}|^2$}
    \put(80.1,28.5){\vtext{\tiny{}$k_x=0.598k_0$}}
    \put(89.5,28.5){\vtext{\tiny{}$k_x=0.602k_0$}}
    \put(0.5,58){\vtext{\footnotesize{}Scattered power}}
    \put(49.5,58){\vtext{\footnotesize{}Scattered power}}
    \put(11.3,26){\vtext{\tiny{}$k_x=-0.598k_0$}}
    \put(18.7,26){\vtext{\tiny{}$k_x=+0.598k_0$}}
    \put(43.5,26){\vtext{\tiny{}$k_x=-0.602k_0$}}
    \put(51.3,26){\vtext{\tiny{}$k_x=+0.602k_0$}}
    \end{overpic}
    \caption{(a) An anti-Brewster angle is engineered at $k_x=0.6k_0$ using $\chia{ee}{xx}=\SI{-4.44e-4}{m}$ and $\chia{ee}{zz}=\SI{6.34e-4}{m}$, predicted using \eqref{eq:antibrewster-zz}. Next, (b-e) shows two solutions of \eqref{eq:antibrewster-quadrupolar}, predicting two anti-Brewster angles. These are plotted in (b), for a case where $S_\text{me}^{yzzx}=-0.285j{\times} 10^{-3}$ m and $\chia{em}{xy}=2.00j{\times} 10^{-5}$ m. Note: $\epsilon_1=1$, $\epsilon_2=2$ and $\lambda_0=c_0/(\SI{300}{THz})$.}
    \label{fig:antibrewster}
\end{figure}

\subsection{Engineered Angular Reflection (``Anti-Brewster'')}
\label{sec:antibrewster}
In addition to suppressing reflection at a particular angle to control the Brewster angle, it is possible to suppress transmission to create what we will call an ``anti-Brewster'' angle. We will consider the same sets of susceptibilities as in Section \ref{sec:brewster}. 
Starting with $\chia{ee}{xx}$ and $\chia{mm}{yy}$ in \eqref{eq:sparams}, setting $|S_{21}|=0$, and solving for the susceptibilities,
\begin{align}
    \chia{ee}{xx} = -\frac{4}{k_0^2\chia{mm}{yy}}\label{eq:antibrewster-tang} \,,
\end{align}
which has no dependence on the angle of incidence ($k_x$, $k_{z,1}$, or $k_{z,2}$). Thus, if \eqref{eq:antibrewster-tang} is satisfied, the metasurface will behave as a mirror with complete reflection, rather than the desired refection at a particular angle of incidence.

Thus, we again consider $\chia{ee}{zz}$, due to its angular behaviour, along with  $\chia{ee}{xx}$. Then, the condition for complete reflection is
\begin{align}
    \chia{ee}{xx} = -\frac{4}{k_x^2\chia{ee}{zz}}\label{eq:antibrewster-zz} \,,
\end{align}
which has a dependence on $k_x$. Designing for $k_x=0.6k_0$, Fig.~\ref{fig:antibrewster}a shows the reflected and transmitted powers, verifying the ``anti-Brewster'' behaviour. 

Finally, we consider quadrupolar susceptibilities. With $S_\text{me}^{yzzx}$ and $\chia{em}{xy}$, there are two solutions for suppressed transmission:
\begin{align}
    \chia{em}{xy} = \frac{\left(2k_x^2-k_0^2\right)S_\text{me}^{yzzx}}{4k_0^2}  \pm \frac{1}{2jk_0}\,,
    \label{eq:antibrewster-quadrupolar}
\end{align}
which are plotted in Fig.~\ref{fig:antibrewster}c and d. The two solutions are very close together, as seen in the magnified plot of Fig.~\ref{fig:antibrewster}e. 
Using the same susceptibilities as in Fig.~\ref{fig:brewster-quad}, the reflected and transmitted powers are plotted in Fig.~\ref{fig:antibrewster}b. We see that the very sharp Brewster angle is straddled by two close ``anti-Brewster'' angles. Thus, this combination of susceptibilities allows for 3 Brewster angles and 2 ``anti-Brewster'' angles.

Overall, we see that by adding more susceptibility terms -- and terms relating to quadrupoles and spatial disperion in particular -- it is possible to have increasing control over the angular scattering response. While we have highlighted a few of the possible terms in the general hypersusceptibity matrix \eqref{eq:spatial-dispersion}, other susceptibilities could be considered for even more intricate control, such as more Brewster or ``anti-Brewster'' angles.


\section{Conclusion}\label{sec:conclusion}
In summary, we have derived GSTCs which include spatial dispersion and are valid for metasurfaces in non-homogeneous environments, such as for practical metasurfaces fabricated on a substrate. We have shown how the susceptiblity tensor properties (symmetries, reciprocity, tracelessness) and spatial symmetries of the metasurface can be used to simplify the susceptibility tensors. Furthermore, we demonstrated how the new hyper-susceptibility terms can be used to produce multiple Brewster and ``anti-Brewster'' angles. For example, with tuning of $\chia{em}{xy}$ and $S_\text{me}^{yzzx}$ it is possible to achieve 3 Brewster angles and 2 ``anti-Brewster'' angles. We expect this work to provide a fundamental advance for Fourier-domain signal processing, where tuning of the angular response is paramount.



\section{Funding}
We gratefully acknowledge funding from the Swiss National Science Foundation (project PZ00P2\_193221).

\section{Supplementary Material}

\subsection{Derivation of quadrupolar GSTCs}
Maxwell's equations in their usual form are only valid for continuous regions and fields. For example, they can be applied inside a dielectric sphere (with appropriate boundary conditions), or outside the sphere, but not in a region which encloses the dielectric interface. Doing so can lead to physical inconsistencies \cite{mithatDiscontinuitiesElectromagneticField2011}. However, Idemen showed that by interpretting Maxwell's equations using distributions, they become valid despite the presence of discontinuities. Considering a discontinuity at $z=0$, all fields quantities $\Lambda(z)$ take the form
\begin{align}
	\Lambda(z)=\{\Lambda(z)\}
	+\sum_{k=0}^\infty\Lambda_k\delta^{(k)}(z),
	\label{eq:generalized-function}
\end{align}
as described in the main text. For example,
\begin{subequations}
	\begin{align}
		\mathbf{D}&=\{\mathbf{D}\} + \sum_{k=0}^\infty \mathbf{D}_k\delta^{(k)}(z)\\
		\overline{\overline{Q}} &=\{\overline{\overline{Q}} \} + \sum_{k=0}^\infty \overline{\overline{Q}}_k\delta^{(k)}(z)\,\,\text{etc...} \,.
	\end{align}
\end{subequations}
However, typically $k$ is limited, and we will only keep $k=0$ for the moment densities, corresponding to a single layer \cite{mithatDiscontinuitiesElectromagneticField2011}. Then, as noted in the main text,
\begin{subequations}
	\begin{align}
		\mathbf{D}&=\epsilon_{0} \{\mathbf{E}\}+\epsilon_{0}\sum_{k=0}^{\infty}\Ev_k\delta^{(k)} + \mathbf{P}_0\delta^{(0)}(z)-\frac{1}{2} \left[\overline{\overline{Q}}_0\delta(z)^{(0)}\right] \cdot \nabla\label{Eq:GSTCDerivationFlux}\\
		\mathbf{B} &= \mu_{0}\left(\{\mathbf{H}\}+\epsilon_{0}\sum_{k=0}^{\infty}\Hv_k\delta^{(k)}+\mathbf{M}_0\delta^{(0)}(z) -\frac{1}{2} \left[\overline{\overline{S}}_0\delta^{(0)}(z)\right] \cdot \nabla\right) \,.
	\end{align}\label{eq:DB-distribution}
\end{subequations}
Now, we will expand the gradient on the right, using the chain rule:
\begin{subequations}
	\begin{align}
		\left[\overline{\overline{Q}}_0\delta(z)\right]  \cdot \nabla &= \left(\overline{\overline{Q}}_0\cdot\nabla_t\right)\delta^{(0)}(z) + 
		\left(\overline{\overline{Q}}_0\cdot\zv\right)\delta^{(1)}(z)\\
		\left[\overline{\overline{S}}_0\delta(z)\right]  \cdot \nabla &= \left(\overline{\overline{S}}_0\cdot\nabla_t\right)\delta^{(0)}(z) + 
		\left(\overline{\overline{S}}_0\cdot\zv\right)\delta^{(1)}(z) \,,
	\end{align}
\end{subequations}
where $\nabla_t=\begin{bmatrix}\partial_x & \partial_y & 0 \end{bmatrix}$ denotes the tangential gradient. Thus, we have
\begin{subequations}
	\begin{align}
		\mathbf{D}&=\epsilon_{0} \{\mathbf{E}\}+(\epsilon_{0}\Ev_0+\mathbf{P}_0)\delta^{(0)}(z)-\frac{1}{2} \left[\left(\overline{\overline{Q}}_0\cdot\nabla_t\right)\delta^{(0)}(z) + 
		\left(\overline{\overline{Q}}_0\cdot\zv\right)\delta^{(1)}(z)\right] \\
		\mathbf{B} &= \mu_{0}\left(\{\mathbf{H}\}+(\Hv_0+\mathbf{M}_0)\delta^{(0)}(z) -\frac{1}{2} \left[\left(\overline{\overline{S}}_0\cdot\nabla_t\right)\delta^{(0)}(z) + 
		\left(\overline{\overline{S}}_0\cdot\zv\right)\delta^{(1)}(z)\right] \right) \,,
	\end{align}\label{eq:DB-distribution-expanded}
\end{subequations}
where we see the the $z$ components of the quadrupolar moments for $k=0$ have changed order to $k=1$ due to the differentiation.

\subsection{Universal Boundary Conditions}
Taking all quantities in Maxwell's equations to be in the form of \eqref{eq:generalized-function}, and separating the different orders, one arrives at a set of \textit{universal boundary conditions}. For a discontinuity at $z=0$, these are expressed \cite{mithatDiscontinuitiesElectromagneticField2011,idemenBoundaryConditionsElectromagnetic1987}
\begin{subequations}
	\begin{align}
		\zv\times\Delta\Hv &= -\nabla\times\Hv_0-j\omega\Dv_0+\Jv_0\\
		\zv\times\Delta\Ev &= -\nabla\times\Ev_0+j\omega\Bv_0\\
		\zv\cdot\Delta\Dv &= \rho_0-\nabla\cdot\Dv_0\\
		\zv\cdot\Delta\Jv &= \rho_0-\nabla\cdot\Dv_0\ \,.
	\end{align}\label{eq:ubc}
\end{subequations}
The singular terms associated with a source ($\Jv_0$ and $\rho_0$) would be known beforehand, but the singular terms in the fields (e.g. $\Ev_0$) are unknown and can be found using \textit{compatibility relations}:
\begin{subequations}
	\begin{align}
		\nabla\times\Hv_k + \nabla\times\Hv_{k-1}
		+j\omega\Dv_k &= \Jv_k\\
		\nabla\times\Ev_k + \nabla\times\Ev_{k-1} - j\omega\Bv_k &= 0\\
		\nabla\cdot\Dv_k + \nabla\cdot\Dv_{k-1} &= \rho_k\\
		\nabla\cdot\Bv_k + \nabla\cdot\Bv_{k-1} &= 0\\
		\nabla\cdot\Jv_k + \nabla\cdot\Jv_{k-1}
		-j\omega\rho_k &= 0 \,,
	\end{align}\label{eq:compatibility-relations}
\end{subequations}
for $k\geq 1$. These provide an infinite set of equations. To make the problem tractable, one assumes that the orders above a certain $k$ are negligible.

\subsection{Iterative substitution into the compatibility relations}
In \eqref{eq:DB-distribution}, we assumed that only $k=0$ is present for the moments, and now we will assume that the terms $k\geq 2$ are zero for the fields. 
Then, re-expressing \eqref{eq:DB-distribution-expanded}, we have
\begin{subequations}
	\begin{gather}
		\Dv_{1,t} =\Pv_{1,t}-\frac{1}{2}\left[\Qt_0\cdot\zv\right]_t  \\
		\Bv_{1,t} =\mu_0\left(\Mv_{1,t}- \frac{1}{2}\left[\St_0\cdot\zv\right]_t\right)\\
		H_{1,z} =\frac{1}{2}\zv\cdot\left[\St_0\cdot\zv\right]-M_{1,z}\\
		E_{1,z} = \frac{1}{\epsilon_0}\left(\frac{1}{2}\zv\cdot\left[\Qt_0\cdot\zv\right]-P_{1,z}\right) \,.
	\end{gather}\label{eq:k2}
\end{subequations}
Substituting \eqref{eq:k2} into \eqref{eq:compatibility-relations} with $k=1$,
\begin{subequations}
	\begin{gather}
		\zv\times\Ev_0 = -\nabla\times\mathbf{E}_1 - j\omega\mathbf{B}_1 
		= \frac{1}{2\epsilon_0}\zv\times\nabla_t\left[\zv\cdot\left(\Qt_0\cdot\zv\right)\right]
		+ \frac{j\omega\mu_0}{2}\left(\St_0\cdot\zv\right)_t\\
		\zv\times\Hv_0 = -\nabla\times\mathbf{H}_1 + j\omega\mathbf{D}_1
		=  \frac{1}{2}\zv\times\nabla_t\left[\zv\cdot\left(\St_0\cdot\zv\right)\right]
		- \frac{j\omega}{2}\left(\Qt_0\cdot\zv\right)_t\\
		\zv\cdot\mathbf{D}_0=-\nabla_t\cdot\mathbf{D}_1=\frac{1}{2}\nabla_t\cdot \left(\Qt_0\cdot\zv\right)_t\\
		\zv\cdot\mathbf{B}_0=-\nabla_t\cdot\mathbf{B}_1=\frac{\mu_0}{2}\nabla_t\cdot \left(\St_0\cdot\zv\right)_t \,,
	\end{gather}\label{eq:k1}
\end{subequations}
where we have made use of the identity $\nabla\times[A(x,y)\zv]=-\zv\times\nabla_t A(x,y)$.

Solving \eqref{eq:k1} for the tangential and normal parts of the fields,
\begin{subequations}
	\begin{gather}
		\Ev_{0,t} = -\zv\times(\zv\times\Ev_0) =  \frac{1}{2\epsilon_0}\nabla_t\left[\zv\cdot\left(\Qt_0\cdot\zv\right)\right]
		- \frac{j\omega\mu_0}{2}\zv\times\left(\St_0\cdot\zv\right)_t\\
		\Hv_{0,t} =-\zv\times(\zv\times\Hv_0)=\frac{1}{2}\nabla_t\left[\zv\cdot\left(\St_0\cdot\zv\right)\right]+\frac{j\omega}{2}\zv\times\left(\Qt_0\cdot\zv\right)_t\\
		D_{0,z} =\frac{1}{2}\nabla_t\cdot \left(\Qt_0\cdot\zv\right)_t\\
		B_{0,z} =\frac{\mu_0}{2}\nabla_t\cdot \left(\St_0\cdot\zv\right)_t \,.
	\end{gather}
\end{subequations}

Thus, from \eqref{eq:DB-distribution-expanded} with $k=0$,
\begin{subequations}
	\begin{multline}
		\Dv_{0,t} =\epsilon_0\Ev_{0,t}+\Pv_{0,t}-\frac{1}{2}\left[\Qt_0\cdot\nabla\right]_t  
		\\=
		\Pv_{0,t}
		+\frac{1}{2}\nabla_t\left[\zv\cdot\left(\Qt_0\cdot\zv\right)\right]
		- \frac{j\omega\mu_0\epsilon_0}{2}\zv\times\left(\St_0\cdot\zv\right)_t
		-\frac{1}{2}\left[\Qt_0\cdot\nabla\right]_t  
	\end{multline}
	\begin{multline}
		\Bv_{0,t} =\mu_0\left(\Hv_{0,t}+\Mv_{0,t}- \frac{1}{2}\left[\St_0\cdot\nabla\right]_t\right)
		\\=
		\mu_0\left(\Mv_{0,t}+\frac{1}{2}\nabla_t\left[\zv\cdot\left(\St_0\cdot\zv\right)\right]+\frac{j\omega}{2}\zv\times\left(\Qt_0\cdot\zv\right)_t- \frac{1}{2}\left[\St_0\cdot\nabla\right]_t\right)
	\end{multline}
	\begin{multline}
		H_{0,z} =\frac{1}{\mu_0}B_{0,z}+\frac{1}{2}\zv\cdot\left[\St_0\cdot\nabla\right]-M_{0,z}
		=\frac{1}{2}\nabla_t\cdot \left(\St_0\cdot\zv\right)_t+\frac{1}{2}\zv\cdot\left[\St_0\cdot\nabla\right]-M_{0,z}\
	\end{multline}
	\begin{multline}
		E_{0,z} = \frac{1}{\epsilon_0}\left(D_{0,z}+\frac{1}{2}\zv\cdot\left[\Qt_0\cdot\nabla\right]-P_{0,z}\right)
		=
		\frac{1}{\epsilon_0}\left(\frac{1}{2}\nabla_t\cdot \left(\Qt_0\cdot\zv\right)_t+\frac{1}{2}\zv\cdot\left[\Qt_0\cdot\nabla\right]-P_{0,z}\right) \,,
	\end{multline}
\end{subequations}
which, with some manipulation, can be re-written
\begin{subequations}
	\begin{gather}
		\Dv_{0,t} =\Pv_{0,t}
		+\frac{1}{2}\nabla_t Q_{0,zz}
		- \frac{j\omega\mu_0\epsilon_0}{2}\zv\times\left(\St_0\cdot\zv\right)_t
		-\frac{1}{2}\left[\Qt_0\cdot\nabla\right]_t  \\
		\Bv_{0,t} =
		\mu_0\left(\Mv_{0,t}+\frac{1}{2}\nabla_tS_{0,zz}+\frac{j\omega}{2}\zv\times\left(\Qt_0\cdot\zv\right)_t- \frac{1}{2}\left[\St_0\cdot\nabla\right]_t\right)\\
		H_{0,z} 
		=\frac{1}{2}(\nabla_t\zv+\zv\nabla):\St_0-M_{0,z}\\
		E_{0,z} =
		\frac{1}{\epsilon_0}\left(\frac{1}{2}(\nabla_t\zv+\zv\nabla):\Qt_0-P_{0,z}\right) \,.
	\end{gather}
\end{subequations}
\subsection{Substitution into the universal boundary conditions}
Finally, we are ready to substitute into the universal boundary conditions, \eqref{eq:ubc}.
\begin{subequations}
	\begin{multline}
		[[\mathbf{z} \times \mathbf{E}]]=\zv\times \nabla_t E_{0,z}-j\omega \mathbf{B}_{0,t}=
		\frac{1}{\epsilon_0}\zv\times \nabla_t\left[
		\frac{1}{2}(\nabla_t\zv+\zv\nabla):\Qt-P_{z}
		\right]\\-j\omega\mu_0\left(\Mv_{t}-\frac{1}{2}\nabla_tS_{zz}+\frac{j\omega}{2}\zv\times\left(\Qt\cdot\zv\right)_t- \frac{1}{2}\left[\St\cdot\nabla\right]_t\right)
	\end{multline}
	\begin{multline}
		[[\mathbf{z} \times \mathbf{H}]]=\zv\times \nabla_t H_{0,z} + j \omega \mathbf{D}_{0,t}=
		\zv\times\nabla_t\left[\frac{1}{2}(\nabla_t\zv+\zv\nabla):\St-M_{z}\right]\\
		+j\omega \left(
		\Pv_{t}
		+\frac{1}{2}\nabla_t Q_{zz}
		- \frac{j\omega\mu_0\epsilon_0}{2}\zv\times\left(\St\cdot\zv\right)_t
		-\frac{1}{2}\left[\Qt\cdot\nabla\right]_t\right) \,.
	\end{multline}\label{eq:gstcs-wrongfactors}
\end{subequations}
Note that we have dropped the `0' subscripts (e.g. $\Qt_0\rightarrow\Qt$) since there is only one term in these series. Given that $\omega^2\mu_0\epsilon_{0}=k_0^2$ and $\omega^2\mu_0=k_0^2/\epsilon_0$, \eqref{eq:gstcs-wrongfactors} can be re-expressed as
\begin{subequations}
	\begin{multline}
		[[\mathbf{z} \times \mathbf{E}]]=
		-j\omega\mu_0\Mv_{t}
		+\frac{k_0^2}{2\epsilon_0}\zv\times\left(\Qt\cdot\zv\right)
		\\
		-\frac{1}{\epsilon_0}\zv\times \nabla_t\left[
		P_{z}-\frac{1}{2}(\nabla_t\zv+\zv\nabla_t):\Qt
		\right]
		+\frac{j\omega\mu_0}{2}\left[
		\left(\St-S_{zz}\tensor{I}\right)\cdot\nabla_t
		\right]_t
	\end{multline}
	\begin{multline}
		[[\mathbf{z} \times \mathbf{H}]]=
		j\omega\Pv_{t}
		+\frac{k_0^2}{2} \zv\times\left(\St\cdot\zv\right)
		-\zv\times\nabla_t\left[M_{z}-\frac{1}{2}(\nabla_t\zv+\zv\nabla_t):\St\right]\\
		-\frac{j\omega}{2}
		\left[
		\left(\Qt-Q_{zz}\tensor{I}\right)\cdot\nabla_t
		\right]_t \,.
	\end{multline}
\end{subequations}
Meanwhile, the normal components are governed by
\begin{subequations}
	\begin{multline}
		\qquad[[\mathbf{z} \cdot \mathbf{D}]]=-\nabla_t\cdot \mathbf{D}_{0} \\
		=-\nabla_t\cdot
		\left(
		\Pv_{t}
		+\frac{1}{2}\nabla_t Q_{zz}
		- \frac{j\omega\mu_0\epsilon_0}{2}\zv\times\left(\St\cdot\zv\right)_t
		-\frac{1}{2}\left[\Qt\cdot\nabla_t\right]_t
		\right)\\
		=-\nabla_t\cdot
		\left(
		\Pv_{t}
		- \frac{j\omega\mu_0\epsilon_0}{2}\zv\times\left(\St\cdot\zv\right)
		-
		\left(\Qt-Q_{zz}\tensor{I}\right)\cdot\nabla_t
		\right)
	\end{multline}
	\begin{multline}
		\qquad[[\mathbf{z} \cdot \mathbf{B}]]=-\nabla_t\cdot \mathbf{B}_{0} \\
		= -\nabla_t\cdot\mu_0\left(\Mv_{t}+\frac{1}{2}\nabla_tS_{zz}+\frac{j\omega}{2}\zv\times\left(\Qt\cdot\zv\right)_t- \frac{1}{2}\left[\St\cdot\nabla_t\right]_t\right)\\
		= -\nabla_t\cdot\mu_0\left(\Mv_{t}
		+\frac{j\omega}{2}\zv\times\left(\Qt\cdot\zv\right)- 
		\left(\St-S_{zz}\tensor{I}\right)\cdot\nabla_t
		\right) \,.
	\end{multline}\label{eq:av-fields}
\end{subequations}

\subsection{Normal field components}
In the main text, it is asserted that the correct form for the average fields is
\begin{subequations}
	\begin{gather}
		\Ev_{\text{av}} =   \Ev_\text{av,t} +\frac{1}{2}\left.\left(\epsilon_1 E_{\text{i},z}+\epsilon_1 E_{\text{r},z}+\epsilon_2 E_{\text{t},z}\right)\right|_{z=0}\zv\label{eq:acting-fields-E}\\
		\Hv_{\text{av}} =  \Hv_\text{av,t} +\frac{1}{2}\left.\left(\mu_1 H_{\text{i},z}+\mu_1 H_{\text{r},z}+\mu_2 H_{\text{t},z}\right)\right|_{z=0}\zv\label{eq:acting-fields-H} \,,
	\end{gather}
\end{subequations}
which we will now show, as was done in \cite{achouriElectromagneticMetasurfacesTheory2021} for the tangential components of the fields.

\begin{figure}[h!]
	\centering
	\begin{overpic}[width=0.5\textwidth,grid=false,trim={0cm 0cm 0cm 0cm},clip,tics=5]{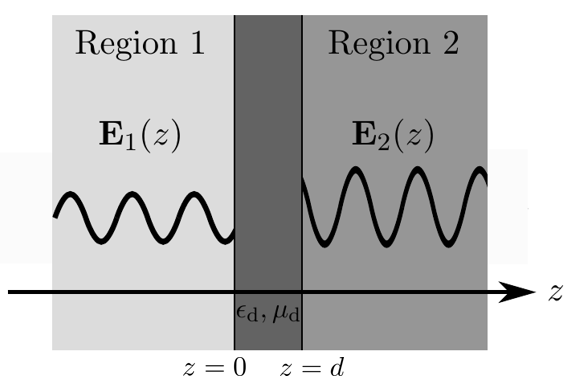}
	\end{overpic}
	\caption{A dielectric slab extending from $0<z<d$ is illuminated by an obliquely incident plane wave. The average electric field within the slab is unknown and desired. This figure is taken from \cite{achouriElectromagneticMetasurfacesTheory2021}.}
	\label{fig:slab}
\end{figure}

Consider a dielectric slab, extending from $0<z<d$ between two different media, and having a relative electric permittivity of $\epsilon_\text{d}$ and a relative magnetic permeability of $\mu_\text{d}$, as in Figure~\ref{fig:slab}. The first medium ($z<0$) is modelled with relataive values $(\epsilon_1,\mu_1)$ while the second medium ($z>d$) is modelled with $(\epsilon_2,\mu_2)$.

Now, with an oblique incident plane wave from the left side (wavevector $\mathbf{k}_\text{d}$ in the slab), the fields within the slab can be expressed as
\begin{align}
	\mathbf{E}_\text{d}(z)=\mathbf{A} e^{-j \beta z}+\mathbf{B} e^{+j \beta z},
\end{align}
where $\beta=\mathbf{k}_\text{d}\cdot\zv$, and constrained by the boundary conditions
\begin{subequations}
	\begin{gather}
		\mathbf{E}_{\text{d},t}(0)=\mathbf{E}_{1,t}(0)  \\
		\mathbf{E}_{\text{d},t}(d)=\mathbf{E}_{2,t}(d)  \\
		\epsilon_\text{d}E_{\text{d},z}(0)=\epsilon_{1} E_{1, z}(0) \\
		\epsilon_\text{d}E_{\text{d},z}(d)=\epsilon_{2} E_{2, z}(d) \,,
	\end{gather}
\end{subequations}
following which one can solve for $\mathbf{A}=\mathbf{A}_{t}+A_z\zv$ and $\mathbf{B}=\mathbf{B}_{t}+B_z\zv$:
\begin{subequations}
	\begin{gather}
		\mathbf{A}_{t}=\frac{\mathbf{E}_{2,t}(d)-\mathbf{E}_{1,t}(0) e^{j \beta d}}{e^{-j \beta d}-e^{j \beta d}} \\
		\mathbf{B}_{t}=-\frac{\mathbf{E}_{2,t}(d)-\mathbf{E}_{1,t}(0) e^{-j \beta d}}{e^{-j \beta d}-e^{j \beta d}}\\
		A_z=\frac{\epsilon_{\mathrm{r}, 2} E_{2, z}(d)-\epsilon_{\mathrm{r}, 1} E_{1, z}(0) e^{j \beta d}}{\epsilon_{\mathrm{r}, \mathrm{d}}\left(e^{-j \beta d}-e^{j \beta d}\right)} \\
		B_z=-\frac{\epsilon_{\mathrm{r}, 2} E_{2, z}(d)-\epsilon_{\mathrm{r}, 1} E_{1, z}(0) e^{-j \beta d}}{\epsilon_{\mathrm{r}, \mathrm{d}}\left(e^{-j \beta d}-e^{j \beta d}\right)} \,.
	\end{gather}
\end{subequations}

Finally, we want to find the averages of these fields. This is done by integrating across the slab and dividing by its thickness,
\begin{subequations}
	\begin{align}
		\mathbf{E}_{\|, \mathrm{av}} & =\frac{1}{d} \int_0^d \mathbf{E}_{\text{d},t}(z) d z \\
		& =\frac{\mathbf{E}_{1,t}(0)+\mathbf{E}_{2,t}(d)}{\beta d} \tan \left(\frac{\beta d}{2}\right) \,,
	\end{align}
	using which we take the average as $d \rightarrow 0$ for a thin surface. Then,
	\begin{align}
		\lim _{d \rightarrow 0} \mathbf{E}_{\text{d},\mathrm{av}}=\frac{\mathbf{E}_{1, t}(0)+\mathbf{E}_{2,t}(0)}{2} \,. \label{eq:t-av}
	\end{align}
\end{subequations}
Meanwhile, the average part of the $z$ component is
\begin{subequations}
	\begin{align}
		E_{z, \mathrm{av}} & =\frac{1}{d} \int_0^d E_{\text{d},z}(z) d z \\
		& =\frac{\epsilon_{1} E_{1,z}(0)+\epsilon_{2} E_{2,z}(d)}{\beta d} \tan \left(\frac{\epsilon_{\mathrm{d}} \beta d}{2}\right) \,.
	\end{align}
	From which
	\begin{align}
		\lim _{d \rightarrow 0} E_{z, \mathrm{av}}=\frac{\epsilon_1E_{1,z}(0)+\epsilon_2E_{2,z}(0)}{2\epsilon_\text{d}} \,. \label{eq:z-av}
	\end{align}
\end{subequations}

Now, for susceptibilities the dimensionless factor in the denominator $\epsilon_\text{d}$ serves as a factor which can be absorbed by the susceptibilities, and so we can set $\epsilon_\text{d}=1$. Thus, noting that $\Ev_1=\Ev_\text{i}+\Ev_\text{r}$ and $\Ev_2=\Ev_\text{t}$, we arrive at \eqref{eq:acting-fields-E}. By duality, the same exercise can be carried out for $\Hv$ fields to find \eqref{eq:acting-fields-H}.


\bibliographystyle{ieeetr}
\bibliography{ville-zotero-1}

\begin{thebibliography}{10}

\bibitem{glybovskiMetasurfacesMicrowavesVisible2016}
S.~B. Glybovski, S.~A. Tretyakov, P.~A. Belov, Y.~S. Kivshar, and C.~R.
  Simovski, ``Metasurfaces: {{From}} microwaves to visible,'' {\em Physics
  Reports}, vol.~634, pp.~1--72, May 2016.

\bibitem{chenReviewMetasurfacesPhysics2016}
H.-T. Chen, A.~J. Taylor, and N.~Yu, ``A review of metasurfaces: Physics and
  applications,'' {\em Rep. Prog. Phys.}, vol.~79, p.~076401, June 2016.

\bibitem{yuFlatOpticsDesigner2014}
N.~Yu and F.~Capasso, ``Flat optics with designer metasurfaces,'' {\em Nat.
  Materials}, vol.~13, Apr. 2014.

\bibitem{genevetHolographicOpticalMetasurfaces2015}
P.~Genevet and F.~Capasso, ``Holographic optical metasurfaces: A review of
  current progress,'' {\em Rep. Prog. Phys.}, vol.~78, no.~2, p.~024401, 2015.

\bibitem{abdollahramezaniMetaopticsSpatialOptical2020}
S.~Abdollahramezani, O.~Hemmatyar, and A.~Adibi, ``Meta-optics for spatial
  optical analog computing,'' {\em Nanophotonics}, vol.~9, pp.~4075--4095, Oct.
  2020.

\bibitem{xueHighNAOpticalEdge2021}
W.~Xue and O.~D. Miller, ``High-{{NA}} optical edge detection via optimized
  multilayer films,'' {\em J. Opt.}, vol.~23, p.~125004, Nov. 2021.

\bibitem{chenOnchipOpticalSpatialdomain2021}
C.~Chen, W.~Qi, Y.~Yu, and X.~Zhang, ``On-chip optical spatial-domain
  integrator based on {{Fourier}} optics and metasurface,'' {\em
  Nanophotonics}, June 2021.

\bibitem{babaeeParallelAnalogComputing2021}
A.~Babaee, A.~Momeni, A.~Abdolali, and R.~Fleury, ``Parallel {{Analog Computing
  Based}} on a \$2\textbackslash ifmmode\textbackslash times\textbackslash
  else\textbackslash texttimes\textbackslash fi\{\}2\$ {{Multiple-Input
  Multiple-Output Metasurface Processor With Asymmetric Response}},'' {\em
  Phys. Rev. Applied}, vol.~15, p.~044015, Apr. 2021.

\bibitem{momeniReciprocalMetasurfacesOnaxis2021}
A.~Momeni, H.~Rajabalipanah, M.~Rahmanzadeh, A.~Abdolali, K.~Achouri,
  V.~Asadchy, and R.~Fleury, ``Reciprocal {{Metasurfaces}} for {{On-axis
  Reflective Optical Computing}},'' {\em IEEE Trans. Antennas Propag.},
  pp.~1--1, 2021.

\bibitem{momeniAsymmetricMetalDielectricMetacylinders2021}
A.~Momeni, M.~Safari, A.~Abdolali, N.~P. Kherani, and R.~Fleury, ``Asymmetric
  {{Metal-Dielectric Metacylinders}} and {{Their Potential Applications From
  Engineering Scattering Patterns}} to {{Spatial Optical Signal Processing}},''
  {\em Phys. Rev. Applied}, vol.~15, p.~034010, Mar. 2021.

\bibitem{achouriAngularScatteringProperties2020}
K.~Achouri and O.~J.~F. Martin, ``Angular {{Scattering Properties}} of
  {{Metasurfaces}},'' {\em IEEE Trans. Antennas Propag.}, vol.~68,
  pp.~432--442, Jan. 2020.

\bibitem{liuGeneralizedBoundaryConditions2019}
X.~Liu, F.~Yang, M.~Li, and S.~Xu, ``Generalized {{Boundary Conditions}} in
  {{Surface Electromagnetics}}: {{Fundamental Theorems}} and {{Surface
  Characterizations}},'' {\em Appl. Sci.}, vol.~9, p.~1891, Jan. 2019.

\bibitem{wangIndependentControlMultiple2020}
X.~Wang, A.~{D{\'i}az-Rubio}, and S.~A. Tretyakov, ``Independent {{Control}} of
  {{Multiple Channels}} in {{Metasurface Devices}},'' {\em Phys. Rev. Applied},
  vol.~14, p.~024089, Aug. 2020.

\bibitem{montiSurfaceImpedanceModeling2020}
A.~Monti, A.~Al{\`u}, A.~Toscano, and F.~Bilotti, ``Surface {{Impedance
  Modeling}} of {{All-Dielectric Metasurfaces}},'' {\em IEEE Trans. Antennas
  Propag.}, vol.~68, pp.~1799--1811, Mar. 2020.

\bibitem{tiukuvaaraSurfaceSusceptibilitiesCharacteristic2022}
V.~Tiukuvaara, T.~J. Smy, K.~Achouri, and S.~Gupta, ``Surface
  {{Susceptibilities}} as {{Characteristic Models}} of {{Reflective
  Metasurfaces}},'' {\em IEEE Transactions on Antennas and Propagation},
  pp.~1--1, 2022.

\bibitem{rahimzadeganComprehensiveMultipolarTheory}
A.~Rahimzadegan, T.~D. Karamanos, R.~Alaee, A.~G. Lamprianidis, D.~Beutel,
  R.~W. Boyd, and C.~Rockstuhl, ``A {{Comprehensive Multipolar Theory}} for
  {{Periodic Metasurfaces}},'' {\em Advanced Optical Materials}, vol.~n/a,
  no.~n/a, p.~2102059, 2022.

\bibitem{dehmollaianComparisonTensorBoundary2019}
M.~Dehmollaian, G.~Lavigne, and C.~Caloz, ``Comparison of {{Tensor Boundary
  Conditions With Generalized Sheet Transition Conditions}},'' {\em IEEE
  Transactions on Antennas and Propagation}, vol.~67, pp.~7396--7406, Dec.
  2019.

\bibitem{zaluskiAnalyticalExperimentalCharacterization2016}
D.~Zalu{\v s}ki, A.~Grbic, and S.~Hrabar, ``Analytical and experimental
  characterization of metasurfaces with normal polarizability,'' {\em Phys.
  Rev. B}, vol.~93, p.~155156, Apr. 2016.

\bibitem{hollowayDiscussionInterpretationCharacterization2009}
C.~L. Holloway, A.~Dienstfrey, E.~F. Kuester, J.~F. O'Hara, A.~K. Azad, and
  A.~J. Taylor, ``A discussion on the interpretation and characterization of
  metafilms/metasurfaces: {{The}} two-dimensional equivalent of
  metamaterials,'' {\em Metamaterials}, vol.~3, pp.~100--112, Oct. 2009.

\bibitem{hollowayHomogenizationTechniqueObtaining2016}
C.~L. Holloway and E.~F. Kuester, ``A {{Homogenization Technique}} for
  {{Obtaining Generalized Sheet Transition Conditions}} ({{GSTCs}}) for a
  {{Metafilm Embedded}} in a {{Magneto-Dielectric Interface}},'' {\em IEEE
  Trans. Antennas Propagat.}, vol.~64, pp.~4671--4686, Nov. 2016.

\bibitem{hollowayCharacterizingMetasurfacesMetafilms2011}
C.~L. Holloway, E.~F. Kuester, and A.~Dienstfrey, ``Characterizing
  {{Metasurfaces}}/{{Metafilms}}: {{The Connection Between Surface
  Susceptibilities}} and {{Effective Material Properties}},'' {\em IEEE
  Antennas Wirel. Propag.}, vol.~10, pp.~1507--1511, 2011.

\bibitem{kuesterAveragedTransitionConditions2003}
E.~Kuester, M.~Mohamed, M.~{Piket-May}, and C.~Holloway, ``Averaged transition
  conditions for electromagnetic fields at a metafilm,'' {\em IEEE Trans.
  Antennas Propag.}, vol.~51, pp.~2641--2651, Oct. 2003.

\bibitem{achouriGeneralMetasurfaceSynthesis2015}
K.~Achouri, M.~A. Salem, and C.~Caloz, ``General {{Metasurface Synthesis
  Based}} on {{Susceptibility Tensors}},'' {\em IEEE Trans. Antennas Propag.},
  vol.~63, pp.~2977--2991, July 2015.

\bibitem{smyFDTDSimulationDispersive2020}
T.~J. Smy, S.~A. Stewart, J.~G.~N. Rahmeier, and S.~Gupta, ``{{FDTD
  Simulation}} of {{Dispersive Metasurfaces With Lorentzian Surface
  Susceptibilities}},'' {\em IEEE Access}, vol.~8, pp.~83027--83040, 2020.

\bibitem{smyIEGSTCMetasurfaceField2021}
T.~J. Smy, V.~Tiukuvaara, and S.~Gupta, ``{{IE-GSTC Metasurface Field Solver}}
  using {{Surface Susceptibility Tensors}} with {{Normal Polarizabilities}},''
  {\em arXiv:2105.05875 [physics]}, May 2021.

\bibitem{smyPartEigenfunctionExpansion2022}
T.~J. Smy and S.~Gupta, ``Part 2 \textendash{} {{Eigenfunction Expansion}}
  ({{EFE}}) {{Analysis}} of {{Cylindrical}} and {{Sectorial Metasurfaces}},''
  July 2022.

\bibitem{bernalarangoUnderpinningHybridizationIntuition2014}
F.~Bernal~Arango, T.~Coenen, and A.~F. Koenderink, ``Underpinning
  {{Hybridization Intuition}} for {{Complex Nanoantennas}} by {{Magnetoelectric
  Quadrupolar Polarizability Retrieval}},'' {\em ACS Photonics}, vol.~1,
  pp.~444--453, May 2014.

\bibitem{nizerrahmeierPartSpatiallyDispersive2022}
J.~G. Nizer~Rahmeier, T.~J. Smy, J.~Dugan, and S.~Gupta, ``Part {{I}}
  -{{Spatially Dispersive Metasurfaces}}: {{Zero Thickness Surface
  Susceptibilities}} \& {{Extended GSTCs}},'' {\em IEEE Transactions on
  Antennas and Propagation}, pp.~1--1, 2022.

\bibitem{achouriExtensionLorentzReciprocity2021}
K.~Achouri and O.~J.~F. Martin, ``Extension of {{Lorentz}} reciprocity and
  {{Poynting}} theorems for spatially dispersive media with quadrupolar
  responses,'' {\em Phys. Rev. B}, vol.~104, p.~165426, Oct. 2021.

\bibitem{achouriMultipolarModelingSpatially2022}
K.~Achouri, V.~Tiukuvaara, and O.~J.~F. Martin, ``Multipolar {{Modeling}} of
  {{Spatially Dispersive Metasurfaces}},'' {\em IEEE Transactions on Antennas
  and Propagation}, pp.~1--1, 2022.

\bibitem{richardsTheoryDistributionsNontechnical1990}
I.~J. Richards and H.~K. Youn, {\em The {{Theory}} of {{Distributions}}: {{A
  Nontechnical Introduction}}}.
\newblock {New York, NY}: {Cambridge University Press}, 1990.

\bibitem{mithatDiscontinuitiesElectromagneticField2011}
I.~Mithat, {\em {Discontinuities in the Electromagnetic Field}}.
\newblock {Wiley-IEEE Press}, 2011.

\bibitem{paniagua-dominguezGeneralizedBrewsterEffect2016}
R.~{Paniagua-Dom{\'i}nguez}, Y.~F. Yu, A.~E. Miroshnichenko, L.~A. Krivitsky,
  Y.~H. Fu, V.~Valuckas, L.~Gonzaga, Y.~T. Toh, A.~Y.~S. Kay, B.~Luk'yanchuk,
  and A.~I. Kuznetsov, ``Generalized {{Brewster}} effect in dielectric
  metasurfaces,'' {\em Nat Commun}, vol.~7, p.~10362, Jan. 2016.

\bibitem{lavigneGeneralizedBrewsterEffect2021}
G.~Lavigne and C.~Caloz, ``Generalized {{Brewster}} effect using bianisotropic
  metasurfaces,'' {\em Opt. Express, OE}, vol.~29, pp.~11361--11370, Mar. 2021.

\bibitem{idemenBoundaryConditionsElectromagnetic1987}
M.~Idemen and A.~H. Serbest, ``Boundary conditions of the electromagnetic
  field,'' {\em Electronics Lett.}, vol.~23, no.~13, pp.~704--705, 1987.

\bibitem{simovskiCompositeMediaWeak2018}
C.~Simovski, {\em Composite {{Media}} with {{Weak Spatial Dispersion}}}.
\newblock {Jenny Stanford Publishing}, 1st edition~ed., Nov. 2018.

\bibitem{albooyehElectromagneticCharacterizationBianisotropic2016}
M.~Albooyeh, S.~Tretyakov, and C.~Simovski, ``Electromagnetic characterization
  of bianisotropic metasurfaces on refractive substrates: {{General}}
  theoretical framework,'' {\em Annalen der Physik}, vol.~528, no.~9-10,
  pp.~721--737, 2016.

\bibitem{achouriElectromagneticMetasurfacesTheory2021}
K.~Achouri and C.~Caloz, {\em Electromagnetic {{Metasurfaces}}: {{Theory}} and
  {{Applications}}}.
\newblock {Hoboken, NJ}: {Wiley-IEEE Press}, 1st edition~ed., May 2021.

\bibitem{kuesterElectromagneticBoundaryProblems2015}
E.~F. Kuester and D.~C. Chang, {\em Electromagnetic {{Boundary Problems}}:
  {{Electromagnetics}}, {{Wireless}}, {{Radar}}, and {{Microwaves}}}.
\newblock {Boca Raton}: {CRC Press}, Oct. 2015.

\bibitem{tretyakovAnalyticalModelingApplied2003}
S.~Tretyakov, {\em Analytical {{Modeling}} in {{Applied Electromagnetics}}}.
\newblock {Norwood, MA, USA}: {Artech House}, 2003.

\bibitem{calozElectromagneticNonreciprocity2018}
C.~Caloz, A.~Al{\`u}, S.~Tretyakov, D.~Sounas, K.~Achouri, and Z.-L.
  {Deck-L{\'e}ger}, ``Electromagnetic {{Nonreciprocity}},'' {\em Physical Rev.
  Appl.}, vol.~10, no.~4, p.~047001, 2018.

\bibitem{riccardiMultipolarExpansionsScattering2022}
M.~Riccardi, A.~Kiselev, K.~Achouri, and O.~J.~F. Martin, ``Multipolar
  expansions for scattering and optical force calculations beyond the long
  wavelength approximation,'' {\em Phys. Rev. B}, vol.~106, p.~115428, Sept.
  2022.

\bibitem{achouriSymmetriesAngularScattering2019}
K.~Achouri and O.~J.~F. Martin, ``Symmetries and {{Angular Scattering
  Properties}} of {{Metasurfaces}},'' in {\em 2019 {{Thirteenth International
  Congress}} on {{Artificial Materials}} for {{Novel Wave Phenomena}}
  ({{Metamaterials}})}, pp.~X--007--X--009, Sept. 2019.

\bibitem{achouriSpatialSymmetriesMultipolar2022}
K.~Achouri, V.~Tiukuvaara, and O.~J.~F. Martin, ``Spatial {{Symmetries}} in
  {{Multipolar Metasurfaces}}: {{From Asymmetric Angular Transmittance}} to
  {{Multipolar Extrinsic Chirality}},'' Aug. 2022.

\bibitem{balanisAdvancedEngineeringElectromagnetics2012}
C.~A. Balanis, {\em Advanced {{Engineering Electromagnetics}}}.
\newblock {Hoboken, USA}: {John Wiley \& Sons}, 2012.

\end{thebibliography}






\end{document}